\documentclass{mcmthesis}
\mcmsetup{CTeX = false,   
        tcn =Intelligent Online Food Delivery System: A Dynamic Model to Generate Delivery Strategy and Tip Advice, problem = {Hanyi Luo, Mengzhan Liufu, Dongrong Li},
        sheet = true, titleinsheet = true, keywordsinsheet = true,
        titlepage = true}
\usepackage{palatino}
\usepackage{mwe}
\usepackage{graphicx}
\usepackage{subcaption}
\usepackage{float}
\usepackage{amsmath}
\usepackage{multirow}
\usepackage{indentfirst}
\usepackage{gensymb}
\usepackage{supertabular}
\usepackage[ruled,lined,commentsnumbered]{algorithm2e}
\usepackage{geometry}

\begin{document}
\setlength{\parindent}{2em}
\linespread{1.75} 
\setlength{\parskip}{0.7\baselineskip} 
\title{}
	\begin{abstract}
	
	Due to the rapid development of online food ordering platforms and rocketing growth of demand, the market is about to saturate soon, and the future trend is to seek efficient utilization of resources. Specifically speaking, food company must have a reliable algorithm to help them produce efficient delivery strategies; individual customers need planning for their decision making in this field. For example, when customers add tip to their order with the sake of controlling or reducing latency. However, few customers know how much tip is enough to reach their desired latency. Therefore, in our paper, we establish a dynamic model to generate delivery strategy for companies and tip advice for customers. We believe that the system we design is more efficient than the currently primitive system.
 
 We first identify the representative buildings and roads from a map of Tianhe District, Guangzhou; Then, we calculate the deliverymen amount in this area based on Regional Gross Domestic Production (regional GDP). At last, a questionnaire is designed to gather original data, which will be expanded later, for simulation. 
 
 Then, we simulate the delivery process and generate delivery strategies using genetic annealing because it can approach a near optimal solution. High-quality delivery queue ensures that those orders can be delivered within an acceptable amount of time. 
 
 Next, we construct regressions to find out relationships between multiple factors and latency and then generate the advisory amount of tip. In extreme cases where there are too many orders or too few deliverymen, some orders can never be delivered on time. Therefore, we use logistic regression to remove those orders. Then we utilize the remaining data to construct the linear regression. We choose linear regression because of its high explainability and the fact that most economic patterns are linear. The regression result is an linear equation which can directly calculate latency with data on several variables, which are identified to be highly-related with latency. Finally, we plug in those values and desired waiting time, getting the advisory tip price. Multiple indexes suggest that our regression results are accurate and reliable.
	
	Some of strengths of our model are accuracy and realistic, time flexible and simple, applicable, economically desirable and interdependent and separable.	
 	
\begin{keywords} 
Intelligent Online Food Delivery System, Genetic Annealing, Advisory Tips, Logistic Regression, Linear Regression
\end{keywords}

\end{abstract}

\medskip

\begin{center}{\Large{Intelligent Online Food Delivery System: A Dynamic Model to Generate Delivery Strategy and Tip Advice}}\end{center}

\begin{center}{\large{Hanyi Luo, Mengzhan Liufu, Dongrong Li}\footnote{International Department, The Affiliated High School of South China Normal University, Guangzhou 510630.}}\end{center}

\section{Abstract}
	Due to the rapid development of online food ordering platforms and rocketing growth of demand, the market is about to saturate soon, and the future trend is to seek efficient utilization of resources. Specifically speaking, food company must have a reliable algorithm to help them produce efficient delivery strategies; individual customers need planning for their decision making in this field. For example, when customers add tip to their order with the sake of controlling or reducing latency. However, few customers know how much tip is enough to reach their desired latency. Therefore, in our paper, we establish a dynamic model to generate delivery strategy for companies and tip advice for customers. We believe that the system we design is more efficient than the currently primitive system.
 
 We first identify the representative buildings and roads from a map of Tianhe District, Guangzhou; Then, we calculate the deliverymen amount in this area based on Regional Gross Domestic Production (regional GDP). At last, a questionnaire is designed to gather original data, which will be expanded later, for simulation. 
 
 Then, we simulate the delivery process and generate delivery strategies using genetic annealing because it can approach a near optimal solution. High-quality delivery queue ensures that those orders can be delivered within an acceptable amount of time. 
 
 Next, we construct regressions to find out relationships between multiple factors and latency and then generate the advisory amount of tip. In extreme cases where there are too many orders or too few deliverymen, some orders can never be delivered on time. Therefore, we use logistic regression to remove those orders. Then we utilize the remaining data to construct the linear regression. We choose linear regression because of its high explainability and the fact that most economic patterns are linear. The regression result is an linear equation which can directly calculate latency with data on several variables, which are identified to be highly-related with latency. Finally, we plug in those values and desired waiting time, getting the advisory tip price. Multiple indexes suggest that our regression results are accurate and reliable.
	
	Some of strengths of our model are accuracy and realistic, time flexible and simple, applicable, economically desirable and interdependent and separable.
	
	\begin{keywords} 
Intelligent Online Food Delivery System, Genetic Annealing, Advisory Tips, Logistic Regression, Linear Regression
\end{keywords}

\tableofcontents

\newpage
\section{Research Significance of the Topic}
After five years of rapid growth and development, online food ordering and delivery industry has already secured its stability. Data$^{[1]}$ shows that the size of the industry in China is growing more than 15\% percent annually. There were 0.26 billion people who use online platforms to order their meals, and forecast shows that the number will approach 0.37 billion in 2020. In 2017, aggregate trade amount on all online platforms exceed 200 billion RMB, and the number will probably escalate into 375 billion in 2018. Nevertheless, it is also undeniable that the growth rate has dwindled when compared with past years’ record: in 2016 the growth rate reach an astonishing 71.3\%, but in 2018 it shrinks to mere 23.5\%.

Such evidence reveals that even platforms such as Meituan, Eleme are still popular, the period of exponential growth has terminated. In large cities, capacity and extent of this industry has reached its maximum; in smaller cities, the marginal cost to develop such large-scale delivery system is astronomical and unaffordable. Moreover, the government recognized the increasingly prominent role of online food ordering pattern, so stricter regulations and policies appeared to constitute deterrent to previously unlimited expansion. 

When the growth remain mild and stable, defects of the system become obvious. Problems such as long waiting time$^{[2]}$, high delivery cost and inefficient assignment of orders start to detract from the industry’s convenience. Under the broad picture, online food ordering platform must endeavor to enhance their systems to retain current prosperity, employing advanced software to efficiently regulate the operation over a large region and sharpening functions to improve customer’s feeling.

Addressing the problems in the online delivery industry, we try to construct a model for better delivery strategy and to predict the tipping amount that the customer should pay in order to get the delivery in certain time.

\section{Introductions and Background}
\subsection{Background}

Entering the application in your mobile phone, choosing your favorite restaurant to order the food and waiting comfortably for the deliveryman, such a pattern to have a meal has grown ubiquitous in China. With the rapid growth of takeout food industry, scholars have paid special attention and emphasis on its economic pattern. Despite the recognition, current system remains primitive, and was lagging behind the fast social development. For example, \textit{the traveling salesman problem (TSP)} mainly focuses on the selling and delivering side, attempting to minimize cost, time etc. for a single delivery route. However, even we are able to precisely design travel route for every deliveryman, we can hardly produce remarkable improvement for the whole market.

Here, we aim to construct a model to consider the system holistically and minimize cost and deadweight loss, seeking the socially optimal point. Our inspiration comes from an interesting phenomenon. Customers can pay certain amount of additional tip. Such an action is to ensure that there are deliverymen to deliver their meal, or to get their meal within a certain latency (waiting time). Unfortunately, the current takeaway food system does not provide an advice for the optimal tip amount to add, which is the least amount of tip that can motivate deliverymen to meet the requirement Economically speaking, the socially optimal amount should be the point where demand and supply meet; neither more nor less is economically optimal. And integrate to the calculation process, we assign each order to the deliveryman who can delivers with minimum cost and find out best travel route for each deliveryman, approaching socially optimal condition.

\subsection{Restatement}

	We are going to build a model of delivery strategy to simulate the delivery process and the tip giving situation as accurate. Hence, the accurate data is an important part. 
	
	First, we need to construct a simulated area with restaurants, destination buildings and simulated orders. To achieve that, we will collect data on number of deliverymen, order distribution and information, population of each destination and latest sales amount of each restaurant. 
	
	Then, based on the data, we will do several simulations. In simulated results, we have the advised route for deliverymen, suggestive tips, latency (waiting time), average tip price, deliverymen number, distance as factors. Ultimately, regression process enables us to generate our final function, which builds a relation between latency and tip price. 
	
	At last, we will analyze the simulation results in general and make conclusions about the methods and problem.
	
\newpage

\section{Assumptions, Justifications and Variables}
\subsection{Assumptions and Justifications}

\noindent\textbf{Assumption 1:} The deliverymen can travel on the roads we denoted in the graph from both directions.\\
\textbf{Justification 1:} The common vehicle for delivery is electrical bicycle. For this type of bicycle, there should be no directional limitation on each side of the road.

\noindent\textbf{Assumption 2:} We only consider delivery time as a customer's latency in our model, which do not include the preparing time from a restaurant.\\
\textbf{Justification 2:} According to real-life experience, restaurants often prepare excess packs of meals beforehand to ensure that the meal can be delivered as soon as possible. What’s more, the tip we discussed mainly acts as motivation for deliverymen, so the preparation time in restaurant is not considered.

\noindent\textbf{Assumption 3:} We don’t include accidents into our consideration.\\
\textbf{Justification 3:} Traffic and other type of accidents are quite unpredictable. Therefore, it is impossible to find any data concerning accidents. Moreover, the tip helps balance the expectation of customers and cost for deliverymen, and neither of them will know about the accidents before they really take place. So such factor will not be considered by both sides.

\noindent\textbf{Assumption 4:} The delivery company only consider the distance and tip price in their decision making.\\
\textbf{Justification 4:} There are other costs of the delivery process, such as traffic costs from deliverymen. However, other variables can be roughly categorize as two costs: distance and tip price. The cost of distance may include oil and electricity cost, and the tip price can reflect the importance of the order. Therefore, with these two factors, we can evaluate the order.

\subsection{Vairables}
\begin{center}
\bottomcaption{Variables}\label{lab:var}
\tablefirsthead{\hline\multicolumn{1}{|c|}{Variables}&\multicolumn{1}{|c|}{Meanings}\\\hline}
\tablelasttail{\hline}
\tablehead{\hline\multicolumn{2}{|c|}
{\small\sl continued from previous page}\\\hline
\multicolumn{1}{|c|}{Variables}&\multicolumn{1}{|c|}{Meanings}\\\hline}
\tabletail{\hline\multicolumn{2}{|c|}{\small\sl continued on next page}\\\hline}

\begin{supertabular}{|p{0.3\textwidth}|p{0.6\textwidth}|}
$S$ & the weight matrix in the Floyd Algorithm\\
$a_n$& real population in each destination building\\
$b_n$& the latest sales amount of the restaurants\\
$A$& the simulated intensity matrix output based on our model\\
$B$&the intensity matrix given in the problem\\
$n$ & sample size dimension\\
$s$ & the frequency of crossover operation in genetic annealing algorithm\\
$\alpha$ & the cooling rate of the genetic annealing algorithm\\
$cost$ & the cost function of the permutation\\
$weight$& the weight of each order\\
$\theta$& the variables used in logistic regression\\
$J(\theta)$ & the cost function used in logistic regression\\
$X_{1}$ & square of ordering time\\
$X_{2}$ & average tip price\\
$X_{3}$ & the proportion of the order that asked for tip\\
$X_{4}$ & number of deliverymen\\
$X_{5}$ & tip price\\
$X_{6}$ & distance from the restaurant to destination of building\\
$X_7$ & distance/tip price\\

\hline
\end{supertabular}
\end{center}

\newpage

\section{Collecting and Processing the Data}

\subsection{Data Collection}
\subsubsection{Data of Map}
	We select Tianhe District, Guangzhou as our simulation area. To start with, we have to select restaurants and destination buildings for simulation, which should represent the real-life situation. According to the official data provided by Meituan Corporation in their quarterly report, we select ten restaurants with the highest sales as our simulated restaurants. Each of these ten restaurants has a monthly sales more than ten thousand orders. The total order amount in Tianhe in June, 2019 is 188342, and the sum of the ten selected restaurants’ sales is 169704. Showing that the top ten restaurants take up nearly 92\% of the takeaway market share in Tianhe, and this huge percentage also render our simulation largely realistic.$^{[3]}$ Similarly, we also select ten destination buildings with largest order records, including South China Normal University, South China University of Technology etc. All the destinations are universities and business areas whose intense crowded population can account for their large order amounts. These ten destinations receive in total 172673 packages, and they take up 91.6\% in the district in June. Moreover, when denoted on the map to form a graph for simulation, the destination \& restaurant web covers 88.2\% of the total area in Tianhe. This percentage is slightly lower than the former two, but it can still account for majority.

	Secondly, we mark those all of the twenty buildings we selected on the satellite map provided by Google Corporation. Since some destination such as South China Normal University encompasses several buildings, we choose one building roughly at the center of the area for our model. After marking all the buildings, all the major roads that can possible be traveled in our delivery web is highlighted. With Gaode and Google GPS, length of each road can be obtained, with which we calculate the traveling distance in delivery.

\begin{figure}[htbp]\centering\includegraphics[width=1\textwidth]{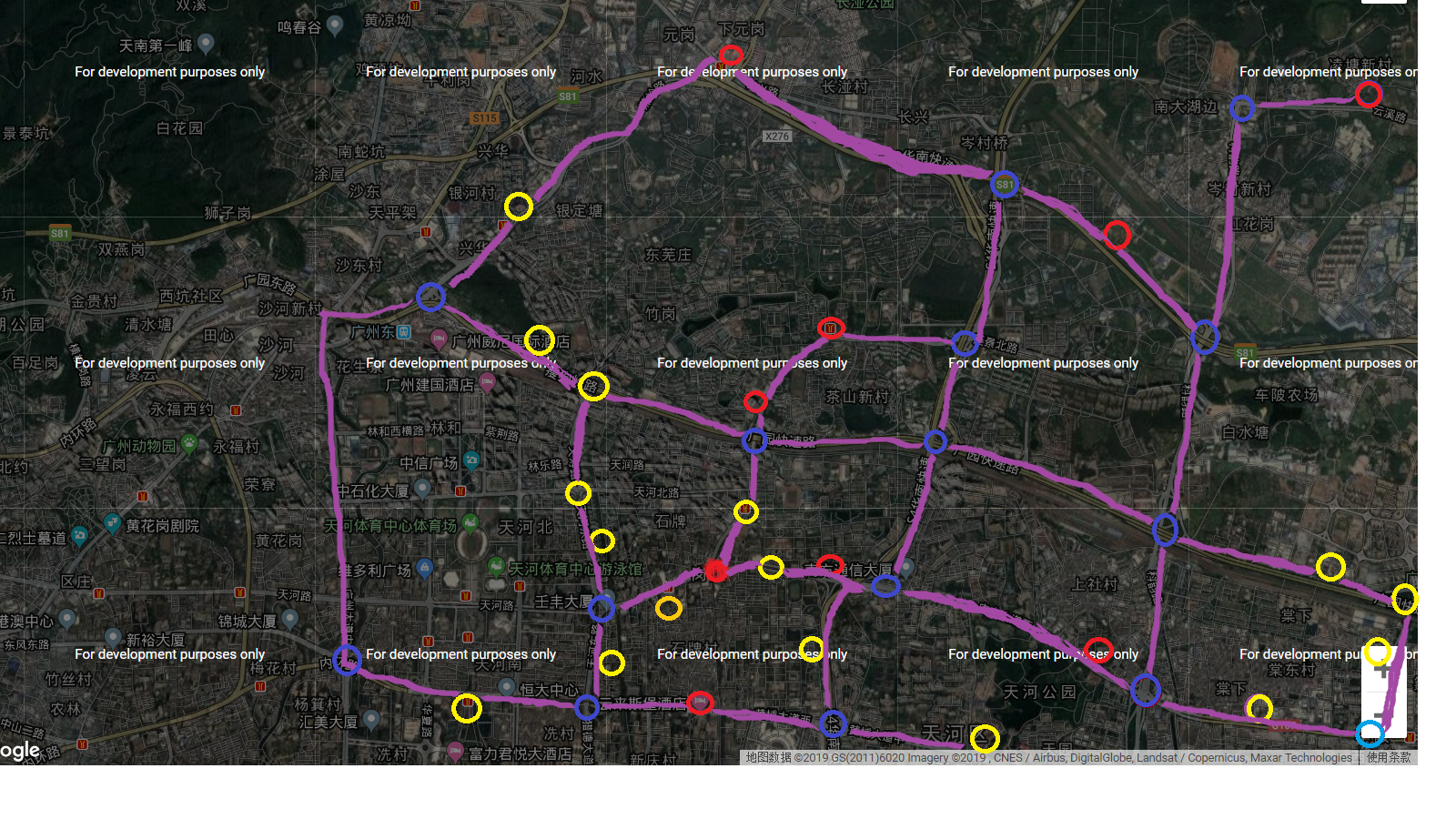}\caption{The Map used to Study} \label{fig:mapp}\end{figure}

The locations of these twenty buildings in total are marked on the satellite map provided by Google Corporation. Each circle in the map represent a vertex in the graph: Restaurants are denoted with yellow circles, destinations are denoted with red circles and crossings in this area are denoted with blue circles. (figure \ref{fig:mapp})

\subsubsection{Estimation of the Number of Deliverymen}
	According to Internet Data Consulting Center$^{[4]}$, the number of deliverymen partly reflect the prosperity of a place. We first identify that the number of deliverymen of a region is correlated with its local gross domestic product (GDP). Therefore, we apply linear regression to the GDP and the number of deliverymen of each cities to measure how the number of deliveryman is associated with the GDP of each cities.

To test whether linear regression is the model of their correlation, we include Pearson Correlation, which can reflect how strongly those two factors are correlated, in our analysis. The Pearson Correlation is obtained by the following formula:
\begin{equation}
\begin{aligned}
r_{xy} \normalsize =\dfrac{\sum_{i=1}^{n} (x_i-\bar{x})(y_i-\bar{y})}{\sqrt{\sum_{i=1}^n(x_i-\bar{x})^{2}}\sqrt{\sum_{i=1}^n (y_i-\bar{y})^{2}}}
\end{aligned}
\end{equation}
in which $r$ has a range between -1 and 1. 

In our calculation, the correlation $r$=0.94709, which means the local GDP is linearly correlated to the number of deliveryman. Hence, calculating the number of deliverymen using linear regression is a suitable method.

 	Linear regression is used to measure the relationship between the two variables. Although there are the inevitable errors, the overall tendency of the relationship can be treated as linear. Hence, we can find correlated factor by the linear fitting.

	Applying the following formula of least square method, we can obtain the fitting function between local GDP and deliverymen:
\begin{equation}
\begin{aligned}\hat{b}=\dfrac{\sum_{i=1}^{n}x_iy_i-n\bar{x}^2}{\sum_{i=1}^{n}x_j^2-n\bar{x}^2}\end{aligned}
\end{equation}
\begin{equation}
\begin{aligned}
\hat{a}=\bar{y}-\hat{b}\bar{x}
\end{aligned}
\end{equation}

	We select several cities and obtained the data of GDP (in 100 million) and rough number of deliverymen $y$ (in thousands) from the internet. 
	
	After calculating, the approximate result comes to: 
\begin{equation}
\begin{aligned}
y=0.0029GDP-10.194
\end{aligned}
\end{equation}

From the Tianhe district official data$^{[5]}$, the annual GDP is 5262.74 hundred million Yuan. Plugging in the formula we derived, the total number of deliveryman is approximately 1516. However, as we mentioned above, our sample area takes up only 82\% of the whole district, the total number of deliverymen should be approximately 1213.

\subsubsection{Data of Order}
	\textbf{Collecting the Raw Data}
	We create a questionnaire on the delivery situation, and received more than 300 responses. With data collected, we are able to simulate the situation of ordering takeout. In this case, we may also learn about the choice distribution for each question, and the distribution is used to expand the data size. The questions and their purposes are as follow:

	First, we asked about the usual waiting time when they order food from their favorite restaurant. The question aimed at finding the delivery distance of the orders. The choice of the question ranges from less than 10 minutes to more than 30 minutes.
 
	To find out the proportion of people who are willing to pay tips to deliverymen, we asked about whether respondent will tip  and how much they would tip. This question can collect the data of the approximate price raising range of customers.	
	
	\textbf{Expanding the Data Scale}
	
Due to the limited capability of our survey, the data size we obtained is not enough for an accurate simulation. Despite the limited size, the ration of each choice on the questions chosen can largely reflect the real situation. Therefore, we devise a proper way to expand our data. 

First, we introduce a larger order group, containing approximately 10257 orders. For each characteristic (delivery time, maximum tip price, delivery distance etc.) that defines a particular order, we find the proportion of a characteristic choice chosen, and the probability for the characteristic to be assigned to an order is equal to the proportion expressed in percentage. For example, if 20\% of the survey orders order food from restaurants within three kilometers, then each order in the larger order group will have a 20\% probability to order food from restaurants within 3 kilometers. We apply this assigning work of every characteristic to every order, then ultimately we procure a larger order group which has roughly the same profile with the survey orders.
	
\subsection{Processing the Raw Data}
\subsubsection{Qualifying the Map}
	\textbf{Qualifying the Structure} We represent the map in reality by using a graph, in which vertices in the graph are the crossroads, restaurants and buildings; the edges are the roads. We give each edge a weight to indicate the length of the road. As we mentioned in the assumption, we suppose all the roads are mutual ways. Therefore, the edges have no direction. 
	
	\textbf{Initializing the Storing Matrix}
	We store the data of the graph by using a weight matrix $S$, in which the element $s_{ij}$ represents the distance between vertex $i$ and $j$ (unit: kilometer). This is an adjacency matrix because the roads are in mutual ways. 
	
	Then we initialize the matrix $S$ by the following rules:
	\begin{enumerate}
	\item If there are direct path from vertex $i$ to $j$, $S_{ij}$ = weight of edge between vertex $i$ and $j$.
	\item If there are no direct path from vertex $i$ to $j$, $S_{ij}$ = $\infty$.
	\item If $i$ = $j$, $S_{ij}$ = 0. (The distance from the vertex to itself is 0.)
	\end{enumerate}
	
\begin{figure}[htbp]\centering\includegraphics[width=1\textwidth]{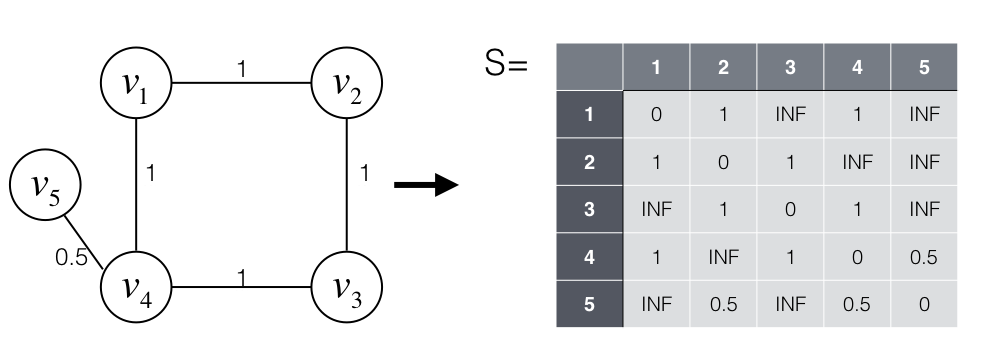}\caption{Qualifying the Map} \label{fig:map1}\end{figure}

	Figure \ref{fig:map1} shows an example of qualifying the map and converting it to an adjacency matrix.

\begin{figure}[htbp]\centering\includegraphics[width=0.8\textwidth]{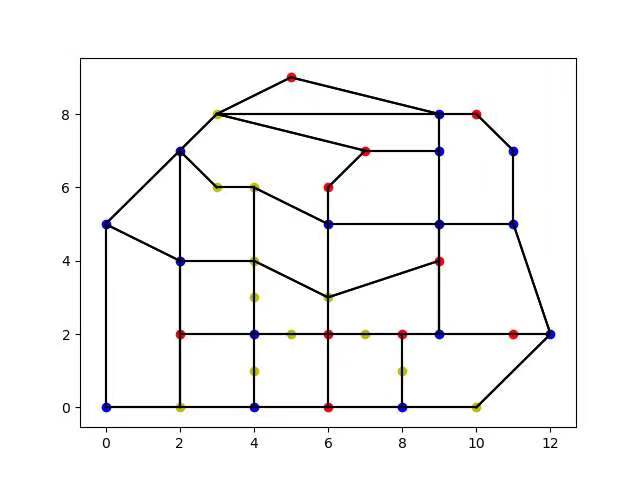}\caption{The Actual Graph} \label{fig:actual}\end{figure}

\subsubsection{Floyd-Warshall Algorithm} 
	To calculate the shortest distance of the graph, we use Floyd-Warshall algorithm. Since we have already initialized the weight matrix $S$, then what the Floyd-Warshall Algorithm do is to change the infinity value into the shortest distance between those two vertices.
	
	The Floyd-Warshall Algorithm is a recursive process. Denote the shortest path between vertex $i$ and $j$ by $Shortest_{ij}$. Let $N$ be the number of vertices in the graph. The general idea of the algorithm is to traversing every path from vertex $i$ to $j$ and find the shortest. Hence, we repeat the following process to find the shortest path:
	\begin{enumerate}
	\item \textbf{renew the distance} Let $k$ = 2. If 
\begin{equation} \begin{aligned} S_{i,k-1}+S_{k-1,j}<S{i,j} \end{aligned}\end{equation}
Then, \begin{equation} \begin{aligned} S_{i,k-1}+S_{k-1,j}=S{i,j}\end{aligned}\end{equation}
	\item \textbf{traversing the situations}
\begin{equation} \begin{aligned} \begin{cases}
 & \text {repeat the steps above},\text{ if } k\leq N \\ 
 & Shortest_{ij}=S_{ij}\text{ if }   k = N
\end{cases} \end{aligned}\end{equation}
\end{enumerate}

	In this way, we are able to calculate the shortest distance from vertex $i$ to $j$. $^{[6]}$ (See appendix)
	
\subsubsection{Order Distribution}
	Since we do not ask the volunteers to indicate specific destinations and restaurants, we must distribute one destination and one restaurant to each simulated order. To achieve that, we note the real population in each destination building with $a$, from $a_1$, $a_2$, ... ,$a_{10}$. Then, the possibility of $P_k$ will be proportional to the population of the building $k$, denoted by $a_k$ versus total population, as expressed as follows:
\begin{equation} \begin{aligned} P_k = \dfrac {a_k}{\sum_{i=1}^{10} a_i}\end{aligned}\end{equation}

Similarly, we denote the latest sales amount of the restaurants we selected with $b$, marking them $b_1,b_2$, ... , $b_{10}$. Since the volunteers indicated the average time interval they receive the food, we first calculate the rough time of the delivery by the Floyd-Warshall Algorithm. The average biking velocity for cyclists is 15.5 km/h (9.6 mph)$^{[7]}$. Hence, the rough time a customer can receive the food can be written as\begin{equation} \begin{aligned} t=\frac{S_{ij}}{15.5} \end{aligned}\end{equation}

We then select those pairs of building within the time constraints as the "candidates", denoted by $C$. 
	
	 The possibility to be assigned to the restaurant with sales amount $b_n$ is
\begin{equation} \begin{aligned} P \text{[be assigned to the restaurant with sales amount $b_k$]} = \dfrac {b_k}{\sum_{i\in C} b_i}\end{aligned}\end{equation}

Then, each of our simulated order has a specific destination building and restaurant.

\subsubsection{Distributing the Deliverymen}
	Distributing by the density of population allows us to relate the density of each area to the deliverymen disperse. Moreover, the pattern of deliverymen dispersing should be approximately close to pattern of the order distribution. We denote the total population of the city by $a_{c}$, and the population of the selected area is $a_{s}$. The rough distribution of the deliverymen by the density of population should be:
\begin{equation} \begin{aligned} P \text{[the deliverymen in the selected area $S$]} = \dfrac{a_{s}}{a_{c}}\end{aligned}\end{equation}

	Since our population density is concentrate on the 10 areas we selected, we discretize the population density into 10 values. Then we distribute the initial location of deliverymen into the 10 most condensed areas.
 
\section{Simulation of Delivery Process}
\begin{figure}[htbp]\centering\includegraphics[width=1\textwidth]{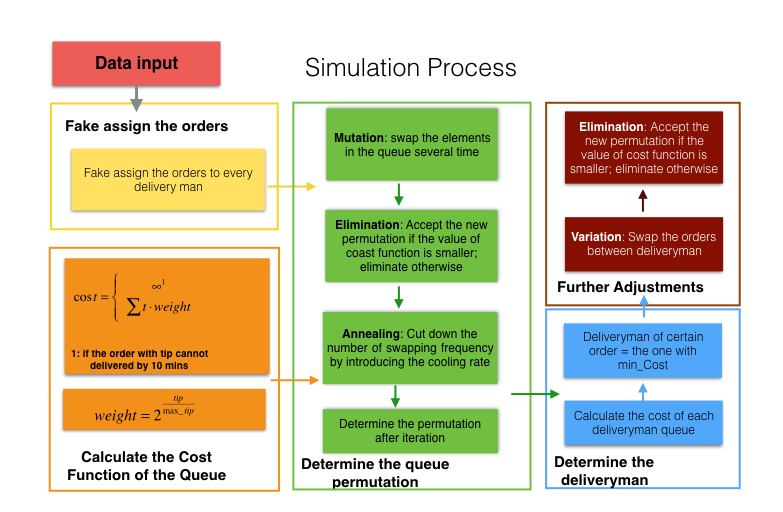}\caption{The Framework to Generate Delivery Strategy} \label{fig:fra}\end{figure}

\subsection{Define the Cost Function}
We define a cost function to determine the goodness of a permutation. The cost function is calculated by the following factors: 

\begin{enumerate}
\item \textbf{Latency:} The most important factor to determine the cost of a permutation is the expected value of latency (waiting time) given the tip price $\mathbb{E}(\text{latency}\mid \text{tip})$. Hence, the cost is positively proportional to the waiting time, which will be high if the waiting time is very long.

\item \textbf{Tip Price:} An order with greater tip price is more important than the order with smaller tip price or without tip.

\item \textbf{Maximum Tip:} In reality, there will be a maximum tip people willing to pay. (i.e. People won’t pay 100 RMB for a single meal delivery tip.) Hence, we set the maximum tip people willing to pay as 100 RMB. (This is a parameter that can be changed based on the society prosperity.)

\item \textbf{Upper Bound for Order with Tips:} An upper bound for $\mathbb{E}(\text{latency}\mid \text{tip})$ is needed to ensure the delivery time.

\end{enumerate}
 
Therefore, the cost function is
\begin{equation}
cost=
\left\{\begin{matrix}
\sum t\cdot  \text{weight}&\\
\infty &\text{if } \mathbb{E}(\text{latency}\mid \text{tip})>20
\end{matrix}\right.
\end{equation}
where,
\begin{equation}
weight=2^{\frac{tip}{max_tip}}
\end{equation}

Hence, the weight is a number between 1 and 2. The order with maximum tips is twice as important as the order without tip. It gives weight to the orders.

\subsection{Genetic Annealing Algorithm}
Genetic annealing algorithm seeks to approach the optimal solution of a problem relying on bio-inspired operators, such as mutation, crossover operation and elimination. $^{[8]}$ Applied to our problem, the algorithm assigns a new order to the deliverymen who minimizes the $\mathbb{E}(\text{latency}\mid \text{tip})$ (the expected value of latency given the tip price). Hence, our goal is to minimize the cost function of the system.

\subsubsection{The Delivery Strategy}
The genetic annealing algorithm generates the high-quality solutions to the permutation of the delivery queue to approach the goal. We apply the algorithm to assign orders to deliverymen by the following steps:

\begin{figure}[htbp]\centering\includegraphics[width=0.8\textwidth]{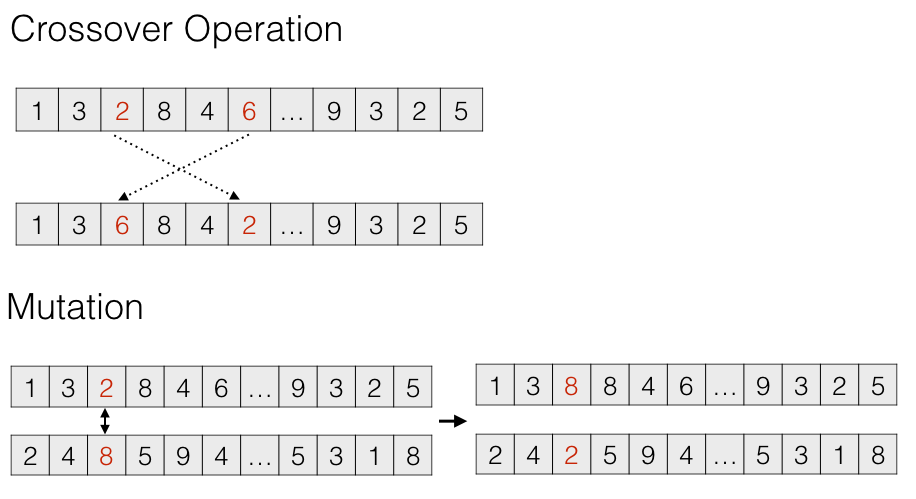}\caption{Examples for Crossover Operation and Mutation} \label{fig:fra}\end{figure}

\textbf{1. Initialization}\\
When we assign a new order to a deliveryman, we first need to determine that the new order assigns to which deliveryman. Hence we "fake assign" the order to every deliveryman and initialize the queue by randomly insert the vertex number into permutation. In this way, we get our first generation. Based on the initial permutation, we calculate the cost function of the initial solution. 
 
\textbf{2. Crossover operation}\\
 We generate another solution by the crossover operation. The basic idea is to modify the chromosomes (which is the permutation of the queue in our context). We regenerate the permutation by randomly swapping two genes in the chromosome (elements in the queue) several times. The initial frequency of the crossover operation is equal to the length of the queue. The process above allows us to get new permutation of the queue for each deliveryman.

 \textbf{3. Selection}\\
 Knowing the first two generations (permutations), we select the solutions by plugging the permutation information into the cost function. We accept the new generation if the cost of the permutation is lower and eliminate otherwise. 
\begin{equation}
\left\{\begin{matrix}
cost_n>cost_{n+1} \; \text{accept new generation}
\\\text{otherwise eliminate new generation}
\end{matrix}\right.
\end{equation} 
This ensures the accepted new generation is always better than the old ones. 
 
\textbf{4. Cooling Schedule}\\
The swapping frequency is 
\begin{equation}
\begin{aligned}
s = \alpha s
\end{aligned}
\end{equation}
in which $\alpha$ is the cooling rate. Since the ideal cooling rate cannot be determined beforehand, we get the value of $\alpha$ after empirically adjusting. Since $\alpha < 1$, step is becoming shorter and shorter correspondingly. This implies that we change the permutation less as the cost becomes lower. The cooling schedule efficiently reduce the running time of our simulation. 

Moreover, since the frequency of the permutation should be an integer, we discrete the value of frequency by the following mechanism. We store a new variable 
\begin{equation}
\begin{aligned}
{s}' = \alpha {s}' 
\end{aligned}
\end{equation}
which changes after every crossover operation. Then we compare the actual swapping frequency $s$ with ${s}'$. If 
\begin{equation}
\begin{aligned}
s-\left \lfloor {{s}'} \right \rfloor>0.9
\end{aligned}
\end{equation}
Then $s={s}'$. By doing this, swapping frequency $s\in \mathbb{Z}$.

\textbf{5. Restarts}\\
The genetic annealing algorithm ends if the cost function value stabilizes, which is the cost function value difference between two generations is zero and at least went through 500 iterations. Otherwise, we restarts from step 1 to step 4 to generate better results.

\textbf{6. Determine the Deliverymen}\\
After the process above, we know the relatively optimal cost function value of every deliveryman within the new order. Therefore, we give the new order to the deliveryman with the lowest cost in the "fake assign" process.

\textbf{7. Further Adjustment: Mutation}\\
Because the result generated above can be improved, we further adjust the permutation by mutation. We generate new solution by swapping the permutation between deliverymen. The cost function value of the system equals the sum of cost of each deliveryman. Then, we accept the solution with lower cost function value of the system. In this way, we further optimize our queue permutation.
\begin{figure}[htbp]\centering\includegraphics[width=1\textwidth]{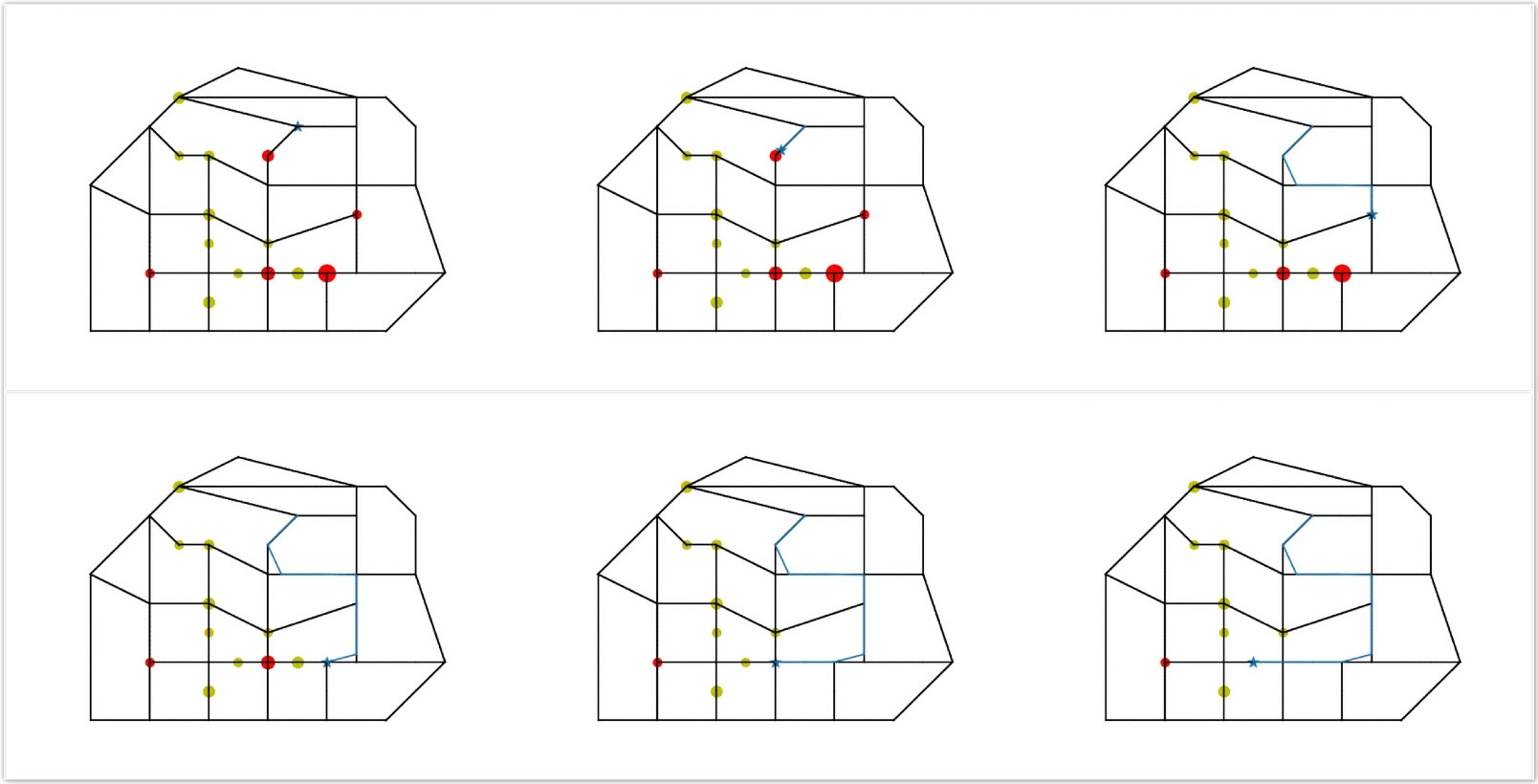}\caption{A Example Route Generated by Genetic Annealing (1)} \label{fig:fra}\end{figure}

\begin{figure}[htbp]\centering\includegraphics[width=1\textwidth]{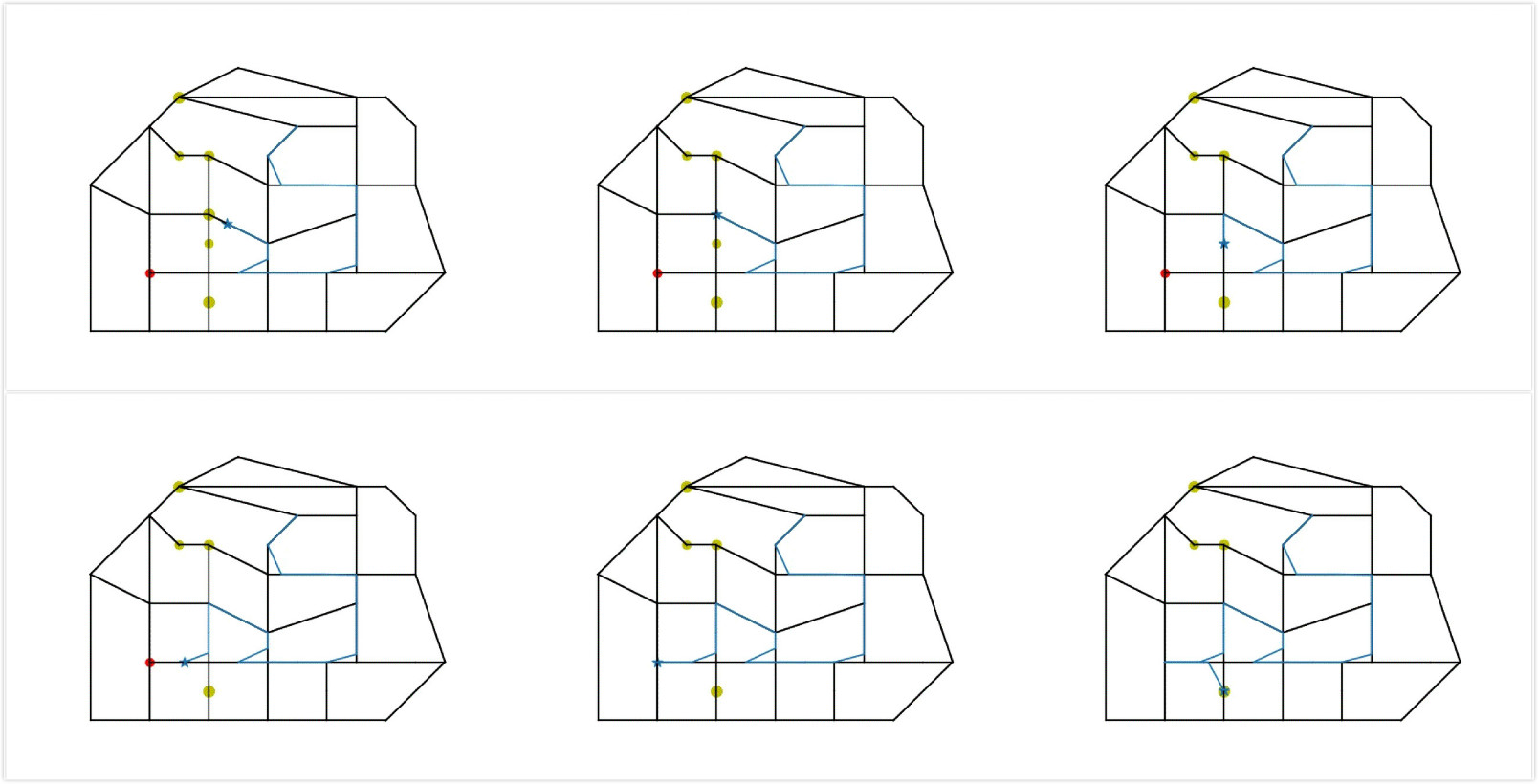}\caption{A Example Route Generated by Genetic Annealing (2)} \label{fig:fra}\end{figure}

\begin{figure}[htbp]\centering\includegraphics[width=1\textwidth]{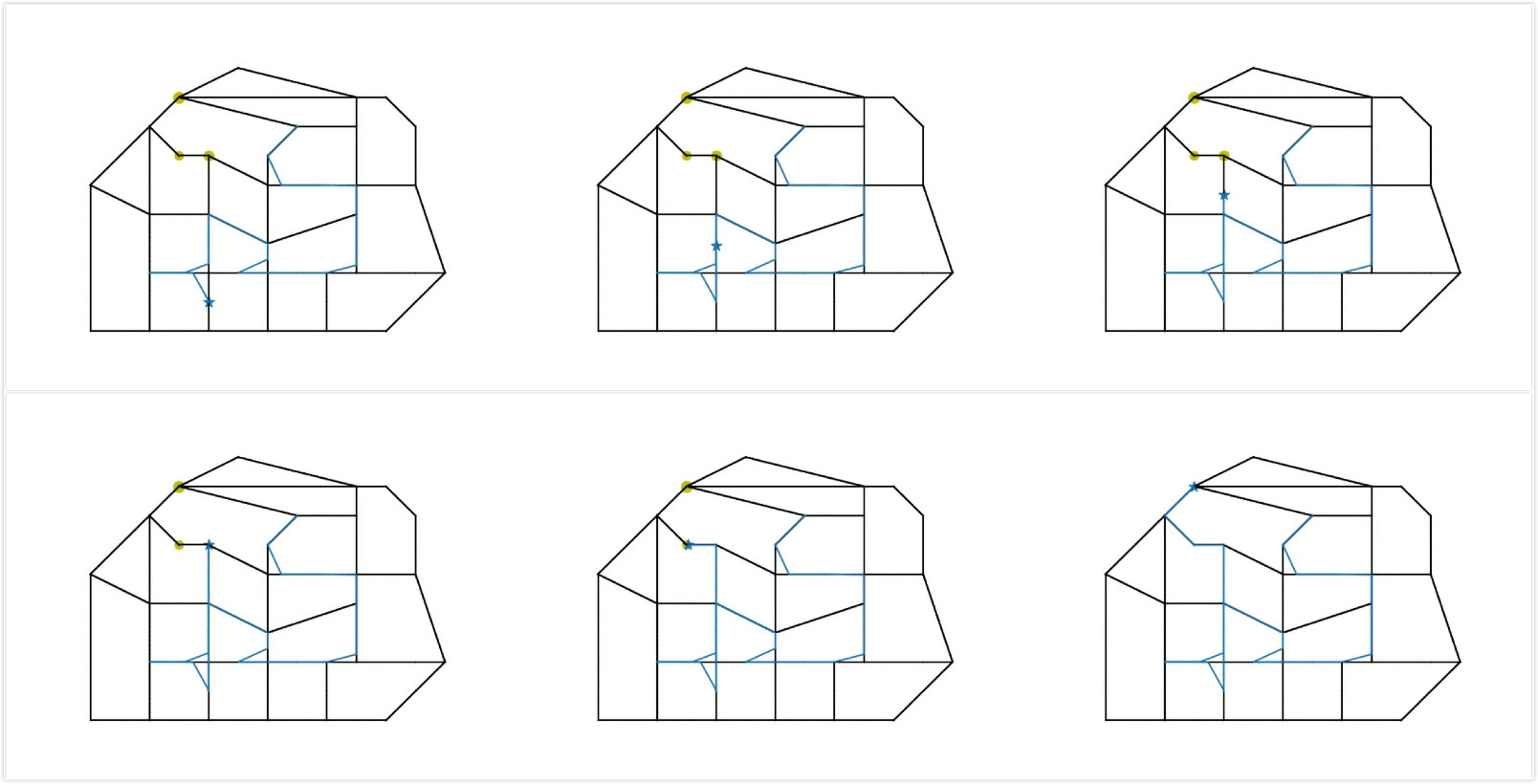}\caption{A Example Route Generated by Genetic Annealing (3)} \label{fig:fra}\end{figure}

\subsection{The Simulation of Real Situation}
We use the genetic annealing to get the high quality delivery strategy for each deliveryman. However, we cannot obtain 100\% real world data. Therefore, in order to simulate the database, which should be as real as possible, we simulate by altering the average tip price and the number of deliverymen. The price is noted according to the normal distribution result of the average price because that conform to the real situation. Since the average price can reflect the local consumption level (reflected by GDP) and the number of delivery man is correlated to the local GDP, those two variables are correlated to each other. We alter these two variables to simulate areas with different consumption level. In this case, we can obtain data that is quite similar to the real situation, and the resultant model can be practical in real situation by inputting the real data.

\begin{figure}[htbp]\centering\includegraphics[width=0.5\textwidth]{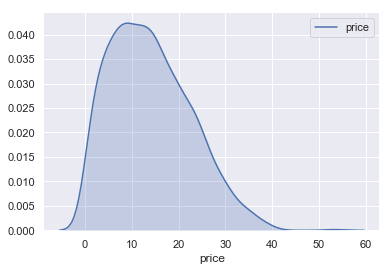}\caption{The Tip Price Distribution} \label{fig:fra}\end{figure}

\subsection{Result Analysis}
\subsubsection{Result Analysis of Genetic Annealing}
\begin{figure}\begin{minipage}[htbp]{0.5\linewidth}\centering\includegraphics[width=1\textwidth]{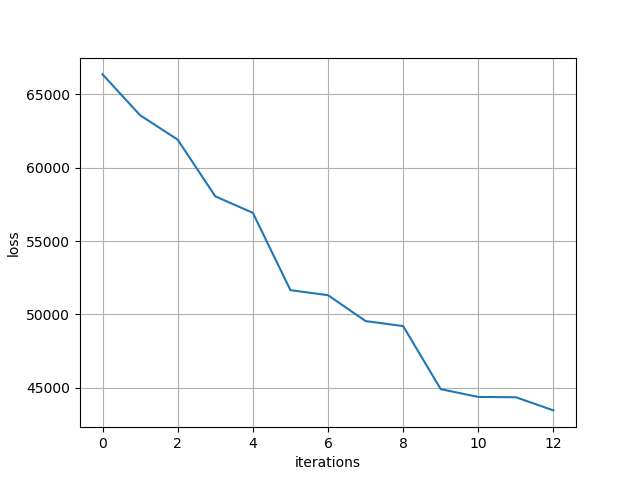}\caption{cost v.s. iteration} \label{fig:result}\end{minipage}\begin{minipage}[htbp]{0.5\textwidth}\centering\includegraphics[width=1\textwidth]{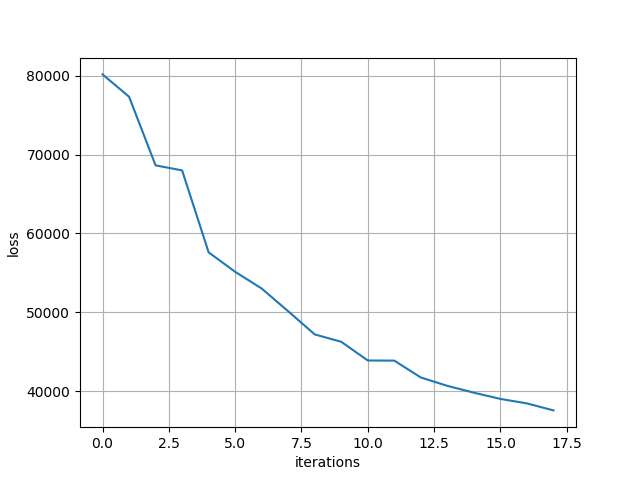}\caption{cost v.s. iteration}\label{fig:cost}\end{minipage}\end{figure}

\begin{figure}\begin{minipage}[htbp]{0.5\linewidth}\centering\includegraphics[width=1\textwidth]{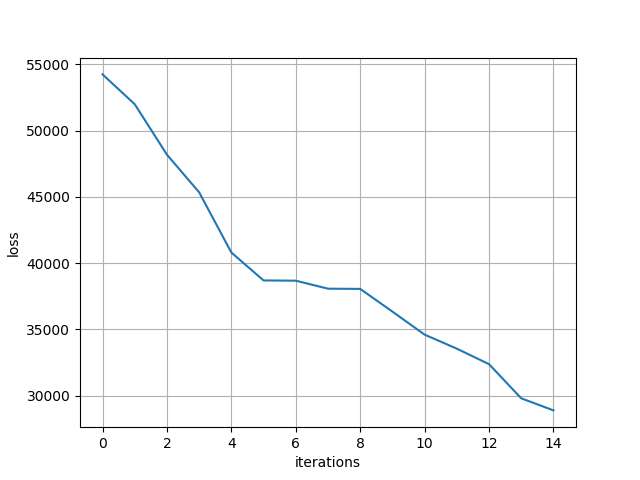}\caption{cost v.s. iteration} \label{fig:result}\end{minipage}\begin{minipage}[htbp]{0.5\textwidth}\centering\includegraphics[width=1\textwidth]{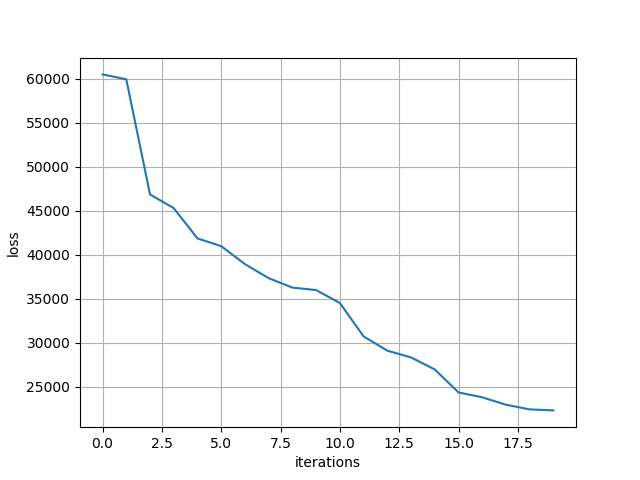}\caption{cost v.s. iteration}\label{fig:original}\end{minipage}\end{figure}

We can use four cost versus iteration graphs (\ref{fig:cost}) to present our operation. We may notice that original cost amount on each graph is different from the other three; they also possess distinct curve shape and terminating points. That’s because genetic annealing is a random algorithm, and it produces different result from every operation. But the four all show decreasing trend of loss, and reach terminating points within 12-18 effective annealing (effective annealing refers to iteration that can lower the loss). This proves that genetic annealing algorithm in our model is able to optimize permutation.

\subsubsection{Result Analysis of the Whole Pattern}
\begin{figure}\begin{minipage}[htbp]{0.5\linewidth}\centering\includegraphics[width=1\textwidth]{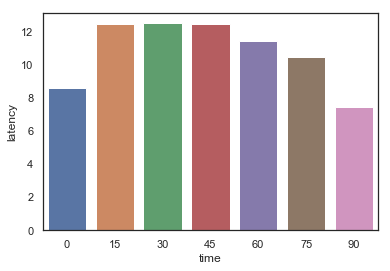}\caption{ordering time v.s. latency} \label{fig:time}\end{minipage}\begin{minipage}[htbp]{0.5\textwidth}\centering\includegraphics[width=1\textwidth]{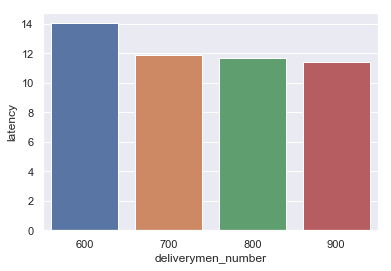}\caption{Deliverymen number v.s. latency}\label{fig:deli}\end{minipage}\end{figure}

The figures presented are the results of our simulation. As we can notice, the pattern indicated in the graphs are reasonable when judged with common sense. The figure \ref{fig:time} (latency vs. Time) shows that the average latency is the smallest at 11:00 and 2:00, which are the start and the end of the time zone we considered in the simulation; in the middle hours the latency is noticeably longer. That is logical: most people order their food, and at two ends the amount should be smaller. And the other figure \ref{fig:deli} that shows the relationship between deliverymen amount and latency demonstrates a simple, reasonable relation: the more deliverymen, the smaller latency. If more workforce is put into a situation with certain workload, more orders can be delivered simultaneously, and the latency is smaller in turn. 

\begin{figure}[htbp]\centering\includegraphics[width=0.6\textwidth]{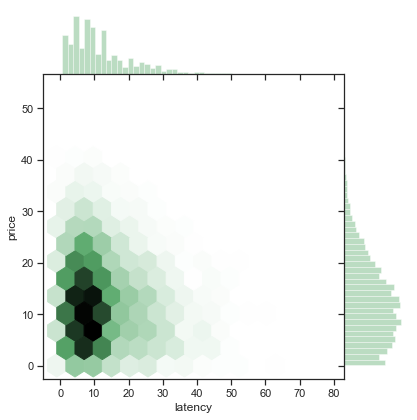}\caption{The density graph of tip price v.s. latency} \label{fig:hetra}\end{figure}

For the density graph \ref{fig:hetra} of tip price versus latency, it also well reflects real-life pattern. The colored portion of the graph appears in the shape of a pyramid. As we can see, if the graph is examined along the vertical axis, at smaller tip price, there are more cases with long latency. For example, if the tip price is zero, there are cases in which latency is approaching forty minutes. By contrast, at tip price at 30 RMB, all orders can be delivered within fifteen minutes. Economically speaking, larger tip price means incentives for deliverymen to prioritize these orders, which would have generally smaller latency. 

\begin{figure}[htbp]\centering\includegraphics[width=1\textwidth]{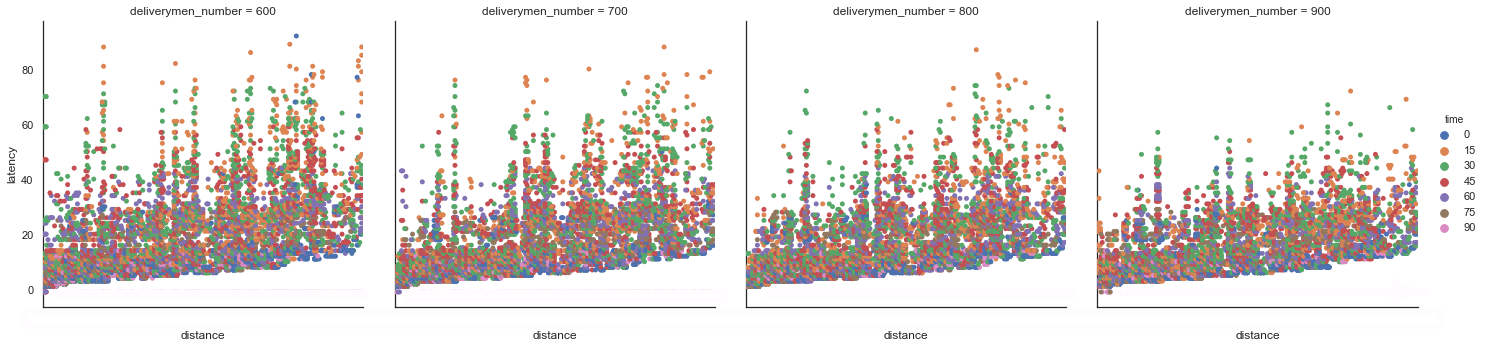}\caption{The Distance v.s. Latency} \label{fig:4}\end{figure}

At last, the four graphs \ref{fig:4} of latency versus delivery distance with specific deliverymen amount are the general presentation of all the orders. When the graph is examined individually, we can observe that latency is positively linear-related to distance, which means if the distance is larger, it would generally take longer to arrive. When the four graphs are compared to each other, we notice that the maximum latency in simulation with nine hundred deliverymen is smaller than those in the rest three graphs, and correspondingly the maximum latency with six hundred deliverymen is the smallest. This pattern corroborates our life experience and inference above.

\newpage

\section{Generate the Advisory Tip}
 \subsection{Logistic Regression}
There are contingencies where deliverymen amount is very small or order amount is very large so some orders cannot be delivered in time (defined as delivery time larger than two hours here), and these particular orders do not contribute to the general relationship between factors. We need a method to determine whether the order can be delivered within two hours or not, and the output of logistic regression (sigmoid function), a widely used classifying algorithm, is binary (either 0 or 1); therefore, logistic regression is a suitable tool here to do the judgments. Orders that can be delivered within two hours are noted with “1”. Those that cannot be are noted with “0”. After the result is obtained, we define a cost function to measure the accuracy of our result. 
 
\subsubsection{Logistic Regression Model}
The output of the logistic regression model would range from 0 to 1, hence 
\begin{equation}
\begin{aligned}
y\in \{0,1\}
\end{aligned}
\end{equation}

\begin{figure}[htbp]\centering\includegraphics[width=0.6\textwidth]{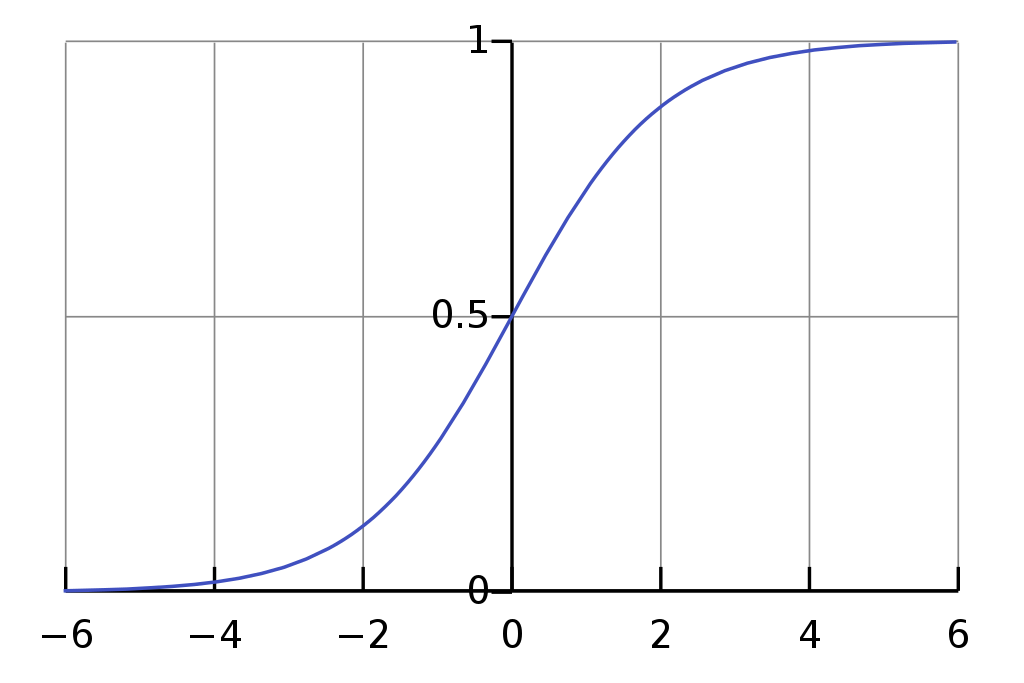}\caption{The Logistic (sigmoid) Function} \label{fig:sig}\end{figure}

\textbf{The sigmoid function}
The general logistic function (the sigmoid function), whose output ranges from 0 to 1, is defined as follows: (See figure \ref{fig:sig})
\begin{equation}
\begin{aligned}
g_\theta(z)=\frac{1}{1+e^{-z}}
\end{aligned}
\end{equation}

\textbf{The general Logistic Function}
Let's assume that $z_\theta(x)$ is a linear function with a series of factors
$x_i$, which are time$^2$, deliverymen number, tip price and average tip price. We can then express $z_\theta(x)$ as follows:
\begin{equation}
\begin{aligned}
z_\theta(x)=\theta^Tx
\end{aligned}
\end{equation}

Then the general logistic function can be modified to
\begin{equation}
\begin{aligned}
p_\theta(x)=g_\theta(z_\theta(x))=\frac{1}{1+e^{-\theta^Tx}}
\end{aligned}
\end{equation}

This measures the probability of the dependent variable $y$ equaling a success (1) rather than a failure (0). $y$ values are not identically distributed: $p(y=1\mid X;\theta)$ differs from one data point $x_{i}$ to another. $^{[9]}$ 

\textbf{Decision Boundary}
After applying the general logistic function, we can determine $y$ by the decision boundary
\begin{equation}
\begin{aligned}
\left\{\begin{matrix}
y=1&\text{if } p_\theta(x)\geq0.5\\ 
y=0&\text{if } p_\theta(x)<0.5
\end{matrix}\right.
\end{aligned}
\end{equation}
which is,
\begin{equation}
\begin{aligned}
p(y;\theta) = p_\theta(x)^y(1-p_\theta(x))^{1-y}
\end{aligned}
\end{equation}

\subsubsection{Cost Function for Logistic Regression}
To find the optimal result of the logistic regression, we define the cost function for the logistic regression with MLE (maximum likelihood estimator) theory. The parameters of our factors $x_i$ in logistic regression which yield minimum value of the cost function will be the optimal parameters. The likelihood function is first built up as follows: 

\begin{equation}
\begin{aligned}
L(\omega) = \prod_{i=1}^{m}p(y;\theta) = \prod_{i=1}^{m}p_\theta(x)^y(1-p_\theta(x))^{1-y}
\end{aligned}
\end{equation}

With the likelihood function, we tend to use the gradient decent to find the minimum (optimal) value. However, it's hard to directly calculate the derivative to get the gradient decent result of the likelihood function in such form. Therefore, we modify the likelihood function in this way and forms the cost function $J(\theta)$
\begin{equation}
\begin{aligned}
J(\theta)=-\frac{1}{m}\sum_{i=1}^{m}[y_i\text{log}(p_\theta(x))+(1-y_i)\text{log}(1-p_\theta(x))]
\end{aligned}
\end{equation}

In this way, our goal is to find the 
\begin{equation}
\begin{aligned}
\arg\min_\theta J(\theta)
\end{aligned}
\end{equation}

We then use the gradient decent to find the solution. The partial derivative of $\theta_j$ is calculated to find the gradient, which is the fastest decreasing direction. Then we descend by step of $\alpha$ in the direction.
 \begin{equation}
\begin{aligned}
 \theta(j):=\theta(j)-\alpha\frac{\partial }{\partial \theta_j}J(\theta)
\end{aligned}
\end{equation}
We repeat the step above until there's no better solutions of the logistic regression (no smaller value of $J(\theta)$). Therefore, we can obtain the result of logistics regression.

\subsubsection{Result Analysis of Logistic Regression}
The regression result is presented below:
\begin{table}[htbp]
\centering
\begin{tabular}{llllll}
\toprule
Factors & Coefficient & t-value & z-value & $P>|z|$ & credible interval\\ 
\midrule
time$^2$	&	0.0154	&	0.002	&	8.476	&	0.00	&	[0.012, 0.019]\\
average tip price	&	0.0276	&	0.001	&	21.145	&	0.00	&[0.025, 0.03]\\
proportion	&	-0.8862	&	0.053	&	-16.778	&	0.00	&[-0.99, -0.783]\\
deliverymen number	&	0.0066	&	3.33E-05	&	196.6	&	0.00	&[0.006, 0.007]\\
tip price	&	-0.0009	&	0.001	&	-1.295	&	0.195	&[-0.002, 0]\\
\bottomrule
\end{tabular}
\caption{Result of Logistic Regression}
\label{tab:weight}
\end{table}

which can also be written as 

\begin{equation}
\begin{aligned}
p_\theta(x)=g_\theta(z_\theta(x))=\frac{1}{1+e^{-(0.0154x_1+0.0276x_2-0.8862x_3+0.0066x_4-0.0009x_5)}}
\end{aligned}
\end{equation}

\begin{figure}[htbp]\centering\includegraphics[width=0.7\textwidth]{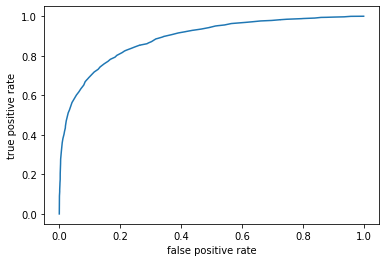}\caption{ROC Curve} \label{fig:roc}\end{figure}

\textbf{Receiver Operating Characteristic}
The Receiver operating characteristic (ROC) is the curve reflect diagnostic ability of logistic regression. Using the composition method, ROC curve reveal the relationship between sensitivity and specificity. It calculates a range of sensitivities and specificities by setting different thresholds for continuous variables. Then, setting the sensitivity as the Y-axis, (1-specificity) as the X-axis, we obtain the following result (figure \ref{fig:roc}). 

To draw an ROC curve, only the true positive rate (TPR) and false positive rate (FPR) are needed (as functions of some classifier parameter). The TPR defines how many correct positive results occur among all positive samples available during the test. FPR, on the other hand, defines how many incorrect positive results occur among all negative samples available during the test.

In this case, the X-axis represents the true positive rate (TPR), which is how many correct positive results occur among all positive samples available during the test. In our context it means how many correct prediction occur among the orders that can be actually delivered on time, and the Y-axis represents the false positive rate (FPR), which defines how many incorrect positive results occur among all negative samples available during the test.$^{[10]}$ Typically, with lower sensitivity comes weaker ability to predict whether the order can be delivered on time. 

The area under the curve (AUC) is equal to the probability that the logistic regression will rank a randomly chosen positive instance higher than a randomly chosen negative one (assuming 'positive' ranks higher than 'negative'). Hence, the larger the area under the curve is, the more precise the chart is. Our graph shows great diagnosis precision, with 0.8909148 AUC. 

 \subsection{Linear Regression}
 
Then the next step is to find the overall relationship between latency and other factors by regression. There are many types of regressions such as linear regression and quadratic regression, but we choose linear regression for our model due to several reasons. First, most economic patterns follow linear relationship (such as the relationship between GDP and deliveryman amount we employed above). Second, linear regression has the best explainability. Once we obtain the result, we try to explain it from logical, everyday-life perspective to interpret the economic pattern underpinning it. For example, the linear curve of latency versus deliveryman amount has a negative gradient, and we interpret it as “More deliverymen, smaller individual workload”. Linear regression best facilitates such interpretation work with its high explainability compared to other regression types. The F-statistics examination and low p-value indicates that linear regression is indeed a suitable model for our case.
 
 \subsubsection{Linear Regression Model}
 
To determine the advisory tip amount, we establish a multi-factor linear regression containing latency, tip amount and other factors. We find all potential variables that can affect the result and do the linear regression to identify the variables used in the regression. Among them, we compare the p-value (which reflects the statistical significance of the factor) and delete the factors which have large p-value. We notice that the p-value for the square of ordering time is smaller than that for time because the rush hours have more orders than other time; therefore, we identify time$^2$ as an effective variable. Finally, we notice that the following variables has the small p-values: average price, ordering time$^2$, deliverymen number, tip price, distance from the restaurant to destination of building, distance/tip price, and proportion (the percentage of tipped-orders and the tip amount).

Since the supposed waiting time is dependent on many variables, we use least square method to optimize linear regression, which is suitable for analyzing multi-factor model. The regression process is shown as follows: 

We consider $Y_{i}$ as the supposed latency, and is the dependent variable of $X_i$, the sextuple independent variable, in each case. $X$ contains 7 components, for $X_{1}$ represents the square of ordering time, $X_{2}$ represents the average tip price, $X_{3}$ represents the proportion of the order that have tip,$X_{4}$ represents the number of deliverymen, $X_{5}$ represents the tip price, $X_{6}$ represents the distance from the restaurant to destination of building, and $X_7$ represents the distance/tip price. $Y_{i}$and $X_{i}$ correlate according to the following pattern:$^{[11]}$ 
\begin{equation}
\begin{aligned}
Y_{i}=\beta _{1}X_{i,1}+...+\beta_{7}X_{i,7}+\varepsilon_{i}
\end{aligned}
\end{equation}
 In which$ Y_{i}, X_{i,1},…, X_{i,7}$ is from the independent identically distributed sample of $Y, X_{1},…, X_{7}$, and $\varepsilon_{i}$ (constant). Let
\begin{equation}
\begin{aligned}
Y=\begin{bmatrix}
Y_{1}\\ 
\vdots\\ 
Y_{n}
\end{bmatrix}
,
X=
\begin{bmatrix}
X_{1,1} & X_{1,2}  & \cdots   & X_{1,7}   \\
X_{2,1} & X_{2,2}  & \cdots   & X_{2,7}  \\
\vdots & \vdots  & \ddots   & \vdots  \\
X_{n,1} & X_{n,2}  & \cdots\  & X_{n,7}  \\
\end{bmatrix}
,
\beta=\begin{bmatrix}
\beta_{1}\\ 
\vdots\\ 
\beta_{7}
\end{bmatrix}
,
\varepsilon=\begin{bmatrix}
\varepsilon_{1}\\ 
\vdots\\ 
\varepsilon_{n}
\end{bmatrix}
\end{aligned}
\end{equation}

Therefore the equation above equivalent to 
\begin{equation}
\begin{aligned}
Y=X\beta+\varepsilon
\end{aligned}
\end{equation}
In which $y_{i}$ and $x_{i}$ are from the data we obtain from simulation process. The goal is to obtain an estimator $b$ of parameter $\beta$. One way to do so is to minimize the following loss function
\begin{equation}
\begin{aligned}
l(\beta)=(Y-X\beta)^{T}(Y-X\beta)
\end{aligned}
\end{equation}
for
\begin{equation}
\begin{aligned}
\frac{\partial }{\partial \beta}(Y-X\beta)^T(Y-X\beta)=-2X^T(Y-X\beta)\\
\frac{\partial^2 }{\partial \beta^2}(Y-X\beta)^T(Y-X\beta)=2X^TX
\end{aligned}
\end{equation}

As long as $X^{T}X$ is positive definite, the estimator $b$ can be written as
\begin{equation}
\begin{aligned}
b=(X^{T}X)^{-1}X^{T}Y
\end{aligned}
\end{equation}
and therefore the linear regression model can be written as $\widehat{Y} =Xb$.

\subsubsection{Regression Coefficient Inference}
In the previous model, $X$ is known, $\beta$ is unknown coefficient, and residual $\epsilon\sim N(0,\sigma^{2}I_{7})$ so
\begin{equation}
\begin{aligned}
b=(X^{T}X)^{-1}X^{T}(X\beta+\epsilon)=\beta+(X^{T}X)^{-1}X^{T}\epsilon
\end{aligned}
\end{equation}
also characterize normal distribution, and 
\begin{equation}
\begin{aligned}
E(b)&=\beta+E[(X^TX)^{-1}X^T\varepsilon]=\beta\\
\text{var}(b)&=\text{var}[(X^{T}X)^{-1}X^T\varepsilon]
		\\&=(X^TX)^{-1}X^T\sigma^2I_{n}X(X^TX)^{-1}
\\&=\sigma^2(X^TX)^{-1}
\end{aligned}
\end{equation}

Since unknown variance exists, we need to construct an estimator of $\sigma^2$. Consider the residual 
\begin{equation}
\begin{aligned}
e&=Y-\widehat{Y}&=X(\beta-b)+\epsilon&= (I_{n}-X(X^{T}X)^{-1}X^T)\epsilon
\end{aligned}
\end{equation}
 noted that 
\begin{equation}
\begin{aligned}
H=X(X^{T}X)^{-1}X^T
\end{aligned}
\end{equation}
Therefore we have
\begin{equation}
\begin{aligned}
E(e)&=(I_{n}-H)E(\epsilon)=0\\
\text{Var}(e)&=(I_{n}-H)^T\sigma^2(I_{7}-H)
&=\sigma^2(I_{7}-H)
\end{aligned}
\end{equation}

Notes that the quadratic sum of the residual as SSE, so
\begin{equation}
\begin{aligned}
SSE=e^{T}e=\varepsilon^{T}(I_{n}-H)\varepsilon
\end{aligned}
\end{equation}

Since $H$ is an idempotent matrix, it has the following property: 
1. rank (A)=tr(A)
2. For $m\times n$ matrix A and $n\times m$ matrix B, tr(AB)=tr(BA).
We can obtain an equation
\begin{equation}
\begin{aligned}
\text{Rank}(H)=tr(X^{T}X(X^{T}X)^{-1})=tr(I_{p})=p
\end{aligned}
\end{equation}
(Where $p$ stands for the column of the matrix $X$.)
Rank$(I_{n}-H)$=tr($I_{n}$)
According to Cochran Theorem, notes that $s^2=SSE/(n-p)$, then $s^2$ is an unbiased estimator. 
Also because SSE and $b$ is relatively independent, so
\begin{equation}
\begin{aligned}
\frac{b-\beta}{s\sqrt{\text{diag}[X^TX]^{-1}}}=\frac{(b-\beta)/\sqrt{\sigma^2\text{diag}[X^TX]^{-1}}}{\sqrt{SSE/(\sigma^2(n-p))}}\sim t(n-p)
\end{aligned}
\end{equation}

\subsubsection{Result Analysis of Linear Regression}
\begin{table}[htbp]
\centering
\begin{tabular}{llllll}
\toprule
Factors & Coefficient & t-value & P>|t| & credible interval\\ 

\midrule
constant	&	-0.1687	&	-0.369	&	0.712	&[-1.066,0.728]\\
time$^2$&	-0.0025&	-20.482	&	0.000	&	[-0.003,-0.002	]\\
average price	&	-0.3371&	-0.369	&	0.713	&[-2.131,1.456]\\
proportion	&	-33.6719	&-3.164	&	0.002	&[-54.537,-12.807]\\
deliverymen number	&	-0.0167&	1.102	&	0.027	&[-0.013,0.046]\\
tip price	&	-0.088	&	-7.383	&	0.000	&[-0.111,-0.065]\\
distance	&	0.0079	&67.478	&	0.000	&	[0.008,0.008]\\
cost (distance/price)	&	0.0010	&	3.254&0.000&	[0.000,0.002]\\
\bottomrule

\end{tabular}
\caption{Result of Linear Regression}
\label{tab:linear}
\end{table}

The result of linear regression is shown in the table \ref{tab:linear}, which can also written as:
\begin{equation}
\begin{aligned}
\text{Latency}=-0.1647-0.0025X_1-0.3371X_2-33.6719X_3-0.0167X_4\\-0.088X_5+0.0079X_6+0.0010X_7
\end{aligned}
\end{equation}

Hence, when the advisory tip price can be generated, if the other information (ordering time, deliverymen number, etc.) is given.

\begin{figure}\begin{minipage}[htbp]{0.5\linewidth}\centering\includegraphics[width=1\textwidth]{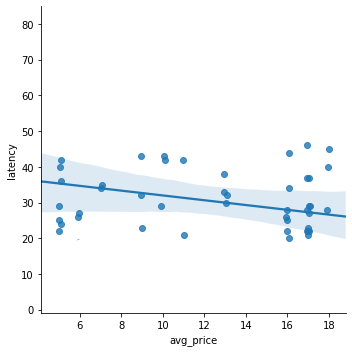}\caption{Average tip price v.s. Latency} \label{fig:lr}\end{minipage}\begin{minipage}[htbp]{0.5\textwidth}\centering\includegraphics[width=1\textwidth]{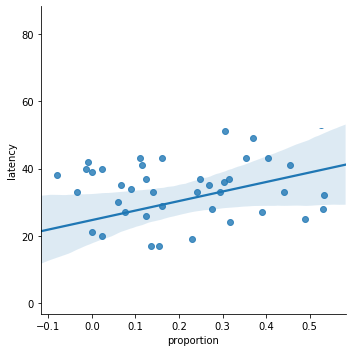}\caption{Proportion v.s. Latency}\label{fig:original}\end{minipage}\end{figure}
 
 \begin{figure}\begin{minipage}[htbp]{0.5\linewidth}\centering\includegraphics[width=1\textwidth]{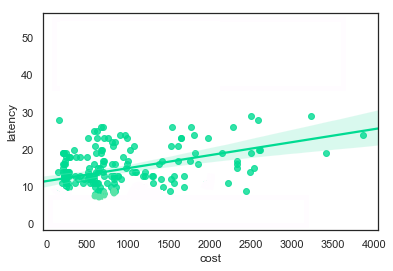}\caption{cost v.s. Latency} \label{fig:result}\end{minipage}\begin{minipage}[htbp]{0.5\textwidth}\centering\includegraphics[width=1\textwidth]{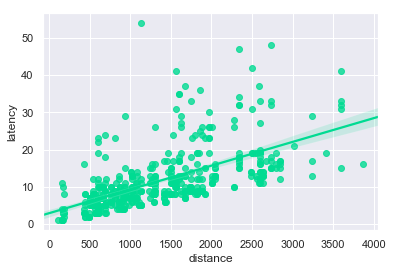}\caption{distance v.s. Latency} \label{fig:result}\end{minipage}\end{figure}
 
 \begin{figure}\begin{minipage}[htbp]{0.5\linewidth}\centering \includegraphics[width=1\textwidth]{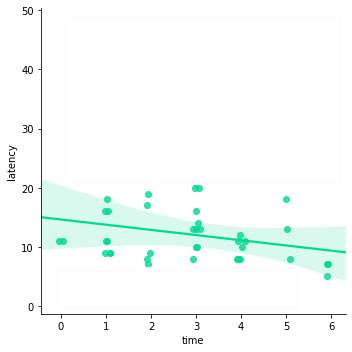}\caption{time v.s. Latency}\label{fig:original}\end{minipage}\begin{minipage}[htbp]{0.5\textwidth}\centering\includegraphics[width=1\textwidth]{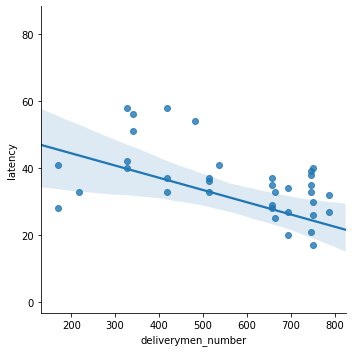}\caption{Deliverymen number v.s. Latency}\label{fig:original}\end{minipage}\end{figure}

By analyzing the six graphs included in figure \ref{fig:lr}, we can examine the accuracy of our regression result. The graphs provide a straightforward presentation of the regression result, and several features prove reliability of the product. Average price in the region, tip price given by customer, proportion of the order that asked for tip, and number of deliverymen are four major factors influencing latency (waiting time). 

The overall F-statistic of the regression is 911.4 and the p-value for the F-statistic is 0.00, which implies the regression produces notable result. Then, for each variable, it owns a very low p-value, demonstrating the factors are effective in the regression.

Shaded areas in the graphs represent confidential interval for each factor identified, and we can see that shaded area only take up relatively small portion in each graph, which means these four factors are highly representative. 

Straight lines drawn in the graph show overall relationship between each factor and latency, and those relationships obtained from regression conform well to real life situation. To demonstrate, it is reasonable to conceive that 
\begin{enumerate}
\item{Deliverymen v.s. Latency}: increment in deliverymen amount will reduce latency, since total order amount will be smaller for individual deliveryman; correspondingly, slope of the straight line in graph latency versus deliverymen number is negative, proving the negatively related relationship.
\item{tip price v.s. latency}: It is definitely true that larger tip price can lift an order’s importance and lift it higher in the deliveryman’s delivery order, in turn cutting latency as suggested by the slope in the graph. 
\item{average price v.s. Latency}: What worth mentioning is that it is harder to understand why average tip price. The average price suggests the local level of consumption. In this way, the place with higher level of consumption tend to add more to the tip price, hence increasing the order's overall importance.
\item{time$^2$ v.s. Latency}: Since time has no straightly linear relationship between the latency, we use time$^2$ to imply the busy hour for delivering. It's reasonable that the latency will be the longest around 12 p.m.. As a result, the slope of the graph is negative. 
\end{enumerate}

\subsection{Further Discussions}
The linear regression is a good method to show the general tip advice within a high efficiency when we apply that in real life. However, the standard deviation of linear regression is not very small, which means the prediction is not perfectly precise. Considering this situation, there are two other more complicated algorithms that can predict the advisory tip with great precision: random forest and XGBoost.

Real life application may involve other factors, such as weather and traffic. The greater the database is, the better precision random forest and XGBoost will have. Therefore, adding these data in calculation may increase the accuracy of decision tree, in turn enhancing the credibility of the two algorithms. However, when in application, fast computation speed is a priority, and including these data into the algorithms make them work far more slowly than regressions. This is also the major reason for our adoption of linear regression.

\subsubsection{Random Forest}
Random forest is a ensemble of several decision trees, which is a set of tree classifiers $\{h(x,\beta_k), k=1,...\}$. Since the output of random forest is discrete, we divide our advisory tip into many small precise intervals. The random forest is constructed by CART algorithm. The basic principle of stochastic forest is to adopt self-help resampling technology. Specifically, the original training sample set $N$ is first determined, and $K$ samples are randomly sampled back and forth to generate a new training set sample set. Then, according to the automatic sample set, a stochastic forest composed of $K$ decision trees is generated. The results of new data are based on the voting result of decision trees.$^{[12]}$ 

In our data, the random forest produces a good accuracy though it's slow computation speed. The accuracy of prediction is 70.28\%, which is a prominent result for our dataset. It will even works better when the data scale gets larger.

\subsubsection{XGBoost}
The idea of the XGBoost (eXtreme Gradient boosting) is to add CART trees and split factors to grow the tree. Each time a tree is added, it is actually to learn a new function to fit the residual of the last prediction.$^{[13]}$  When we get $k$ trees after training, we need to predict the score of a sample. In fact, according to the characteristics of the sample, it will fall to a corresponding leaf node in each tree. Each leaf node corresponds to a score, which is what we only have to add up from each tree to be the advisory tip.
\begin{equation}
\begin{aligned}
\hat{y}=\o (x_i)=\sum_{k=1}^{K}f_k(x_i)
\end{aligned}
\end{equation}
where,
\begin{equation}
\begin{aligned}
F=\{f(x)=w_{q(x)}\}(q:R^m\rightarrow T,w\in R^T)
\end{aligned}
\end{equation}

\begin{figure}[htbp]\centering\includegraphics[width=1\textwidth]{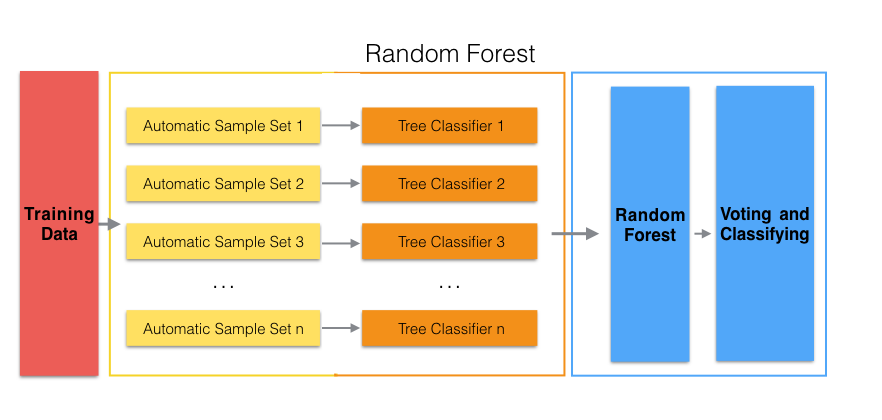}\caption{Random Forest} \label{fig:fra}\end{figure}

\begin{figure}[htbp]\centering\includegraphics[width=0.6\textwidth]{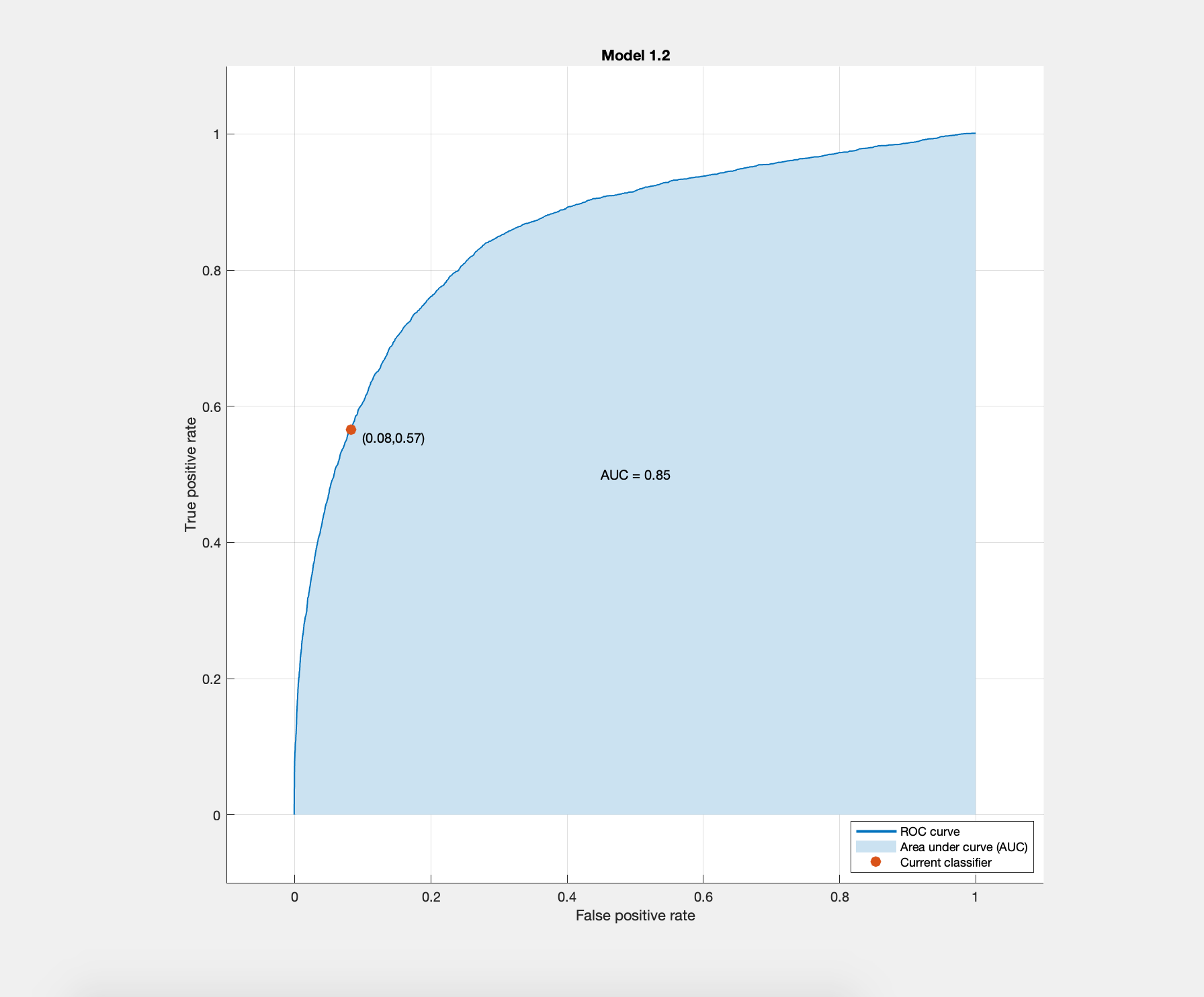}\caption{The AUC of Random Forest} \label{fig:fra}\end{figure}

\section{Strengths and Weaknesses}
\subsection{Strengths}
\begin{enumerate}
\item\textbf{Economically Desirable:} Unlike many typical decision making process, planning models and mechanisms which often base consideration on individual, our model approach a global optimal solution. Not only producing optimal tip price for each customer, our model guarantees that the deliveryman deliver the order with the least cost. Consequently, our model generally produces the optimal delivering strategy for the all orders and deliverymen, reaching socially optimal condition, which is also a long-term goal for companies and economists.

\item\textbf{Applicable:} Our model is devised to tackle delivery problem, but similar commercial platforms such as online-booking taxi, mails etc. Encompassing route planning and order assigning can all employ our model. We can reach such applicability because the factors we considered like distance are universally cost-related; other aspects that are unique to food delivery are not included.  

\item\textbf{Accuracy and Realistic:} We create our graph based on real city location in a district in Guangzhou, and the length of every road in the graph is accurately recorded. Our original data is collected from real respondents, and the expansion procedure is able to appropriately reflect the real-life situation. The accurate data provides a reliable result of the linear regression after simulation process.

In the regression step, all of the factors have low p-value. Several factors reveal that our regression can accurately represent the relationship between the latency and tip price without introducing too much complexity.

\item\textbf{Time Efficiency and Simplicity:} Although the data collection process using simulation seems time-consuming, its application in real life can be quite efficient. Actually, we don’t have to gather data and perform new simulation frequently. Condition over a particular area only fluctuates mildly. Plus, the variation gets smaller when data size grows larger. Once we gather data of a relatively long period such as one month or a few weeks and do simulation, the calculating equation produced is straightforward and can be used for a few weeks or longer. What’s more, once we complete our simulation, regression etc. and obtain the equation to calculate minimum tip, the determination of min tip for each order would be simple.

\item\textbf{Interdependent and Separable:} Our model can be considered as a set of sub-models. Each sub-model can function individually, and can be applied to various problem. Moreover, when we need to modify the model, the corresponding sub-model can be taken out and adjust separately. The adjustment process does not need change the whole model, which is time efficient.

\end{enumerate}

\subsection{Weaknesses}
\begin{enumerate}
\item Special conditions are inevitable in real life, and those accidents, for example, storm and traffic jam, will definitely affect delivery procedure. In order to avoid too much complexity, we neglect those factors in our model.

\item All economical solution should be pragmatic, but our model suggests a system that requires a more refined system, stronger GPS, better communication mechanism between deliveryman and controlling center to illustrate to be implemented. 

\item Our graph and routes included in the model is simpler than real city road webs, and with the sake of maximizing profit, a really large area may be considered as a whole for planning in implementation. That will rise complexity and simulation time in our model.

\item Though $r$ square value in linear regression section is an acceptable result; however, it indicates that our consideration is still far from thorough and perfect. That’s because there are still other factors involving in this situation, such as weather, traffic condition. Due to limited time for model construction and inability to collect data of those factors, we did not achieve a perfect consideration. If better data sources are provided, our model can generate even better result.

\end{enumerate}
\newpage

\section{Conclusion}
According to result analysis above, we conclude that our model is able to generate accurate, helpful proposed tip for minimizing delivery time. With simulation processes and the information of food orders, our model is capable of automatically assigning orders to most suitable deliverymen and instructing them to take suitable routes, and therefore form an optimal delivery strategy. Ultimately, it produces a function to calculate minimum tip price to minimize latency. The function is straightforward and thus efficient in real life application.

Moreover, the pattern and focus of future food delivery industry are demonstrated by our model. The system incorporates all roads, buildings and deliverymen in the region into an integrated system and assign tasks and routes according to precise calculation. Such centralized control is the definite trend, for companies and other economic entity must approach global optimal point from a higher perspective. Unfortunately, the current system, in which deliverymen accept orders themselves, is quite primitive because individual can hardly evaluate cost and design best routine to travel. 
For customer’s point of view, specification of minimum tip price prevents unnecessary expense, resulting in more efficient distribution of capital. In the broadest term, economists insist that society should pursue the socially optimal point, on which aggregate demand and aggregate supply meet; our model contrives to derive minimum cost for the whole system, illustrating the optimal concept well.

\newpage
\section{References}

\noindent [1] “take-outO2O: How much do you know about the data that comes from take-outs.” take-outO2O: How much do you know about the data that comes from take-outs, \\www.woshipm.com, 18 Nov. 2016, www.woshipm.com/data-analysis/446323.html.\\
\medskip
\noindent [2] “The Reason That Deliverymen Are Late.” 22 Feb. 2018, c.m.163.com/news/a/\\E8K7BOKQ0544026J.html?spss=wap\_refluxdl\_2018.\\
\medskip
\noindent [3] Bing, Sun. “MeiTuan 2018 Financial Report.” CEWEEKLY.CN, 12 Mar. 2019, 14:08, www.ceweekly.cn/2019/0312/251399.shtml.\\
\medskip
\noindent [4] DinK. “Internet data consulting center.” 199IT, 21 Jan. 2019, www.199it.com/archives\\/823693.html.\\
\medskip
\noindent [5]  http://www.thnet.gov.cn/thxxw/sjkf/sjkf\_index.shtml\\
\medskip
\noindent [6.1] “Free Algorithms Book.” Free Algorithms Book, May 2018, books.goalkicker.com\\/AlgorithmsBook/.\\
\medskip
\noindent [6.2] Floyd, Robert W. (June 1962). "Algorithm 97: Shortest Path". Communications of the ACM. 5 (6): 345. doi:10.1145/367766.368168\\
\medskip
\noindent [7] Bicycle statistics. City of Copenhagen website. City of Copenhagen. 13 June 2013. Archived from the original on 12 December 2013. Retrieved 12 December 2013.\\
\medskip
\noindent [8.1] Eiben, A. E. et al (1994). "Genetic algorithms with multi-parent recombination". PPSN III: Proceedings of the International Conference on Evolutionary Computation. The Third Conference on Parallel Problem Solving from Nature: 78–87. ISBN 3-540-58484-6.\\
\noindent [8.2] Ting, Chuan-Kang (2005). "On the Mean Convergence Time of Multi-parent Genetic Algorithms Without Selection". Advances in Artificial Life: 403–412. ISBN 978-3-540-28848-0.\\
\medskip
\noindent [9.1] David A. Freedman (2009). Statistical Models: Theory and Practice. Cambridge University Press. p. 128.\\
\noindent [9.2] Rodríguez, G. (2007). Lecture Notes on Generalized Linear Models. pp. Chapter 3, page 45 – via http://data.princeton.edu/wws509/notes/\\
\medskip
\noindent [10.1] /@narkhedesarang. “Understanding AUC - ROC Curve.” Medium, Towards Data Science, 26 May 2019, towardsdatascience.com/understanding-auc-roc-curve-68b23\\03cc9c5.\\
\noindent [10.2] Fawcett, Tom (2006); An introduction to ROC analysis, Pattern Recognition Letters, 27, 861–874.\\
\medskip
\noindent [11.1] Johnson, Richard A. Interfaces, vol. 9, no. 1, 1978, pp. 118–119. JSTOR,\\ www.jstor.org/stable/25059709.\\
\noindent [11.2] Assuncao, Renato, and Paul D. Sampson. Journal of the American Statistical Association, vol. 88, no. 421, 1993, pp. 383–383. JSTOR, www.jstor.org/stable/2290746.\\
\medskip
\noindent [12] Ho TK (1998). "The Random Subspace Method for Constructing Decision Forests" (PDF). IEEE Transactions on Pattern Analysis and Machine Intelligence. 20 (8): 832–844. doi:10.1109/34.709601.\\
\medskip
\noindent [13] Chen, Tianqi; Guestrin, Carlos (2016). "XGBoost: A Scalable Tree Boosting System". In Krishnapuram, Balaji; Shah, Mohak; Smola, Alexander J.; Aggarwal, Charu C.; Shen, Dou; Rastogi, Rajeev (eds.). Proceedings of the 22nd ACM SIGKDD International Conference on Knowledge Discovery and Data Mining, San Francisco, CA, USA, August 13-17, 2016. ACM. pp. 785–794. arXiv:1603.02754\\
\newpage

\section{Acknowledgements}
As most of other high school students, we order takeout food at school. However, due to the fixed schedule of the classes, we sometimes add additional fee in order to make the food arrive at the time we want. We then realized adding additional price could alter the delivery process; therefore we started to conduct research on this topic to generate a economically desirable delivery strategy and tip advice for the takeaway food. 

Guowei Cai is an economics professor at Sun Yat-sen University. He helps us find the general direction of our research and provides us with decent advises on how the research should process with no mercenary relationship. 

Therefore, we wish to express sincere gratitude to professor Guowei Cai, who guided us until this paper is finished. Without the help of our tutor, we would not be able to complete our first academic paper.

We sincerely thank our parents for their support and help in life. They also provide us with valuable suggestions on this essay.

We also thank Yau High School Science Award, for endowing us such a precious opportunity to be involved in economic modeling in high school.

During the writing of this paper, Mengzhan Liufu and Dongrong Li is responsible for mathematical derivation, thesis writing and mathematical modeling. Hanyi Luo is responsible for thesis writing, algorithm design and programming realization.

\section{Appendices}
\subsection{Questionnaire Survey and Results}
\begin{enumerate}
\item What time do you usually order take-away food at noon?
\begin{figure}[htbp]\centering\includegraphics[width=0.7\textwidth]{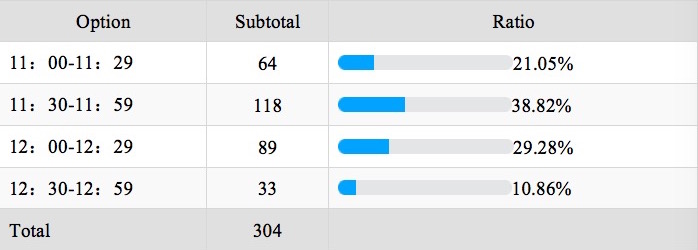}\caption{The answer result for question 1} \label{fig:fra}\end{figure}
\item How long is your waiting time for take-away order?
\begin{figure}[htbp]\centering\includegraphics[width=0.7\textwidth]{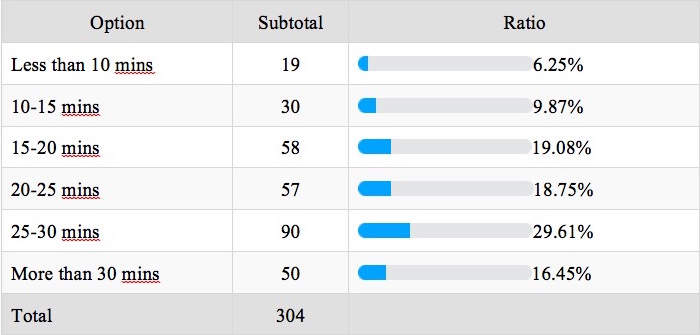}\caption{The answer result for question 2} \label{fig:fra}\end{figure}
\item If it's raining today, will you give tips to deliverymen of the take-away in order to get the food faster?
\begin{figure}[htbp]\centering\includegraphics[width=0.7\textwidth]{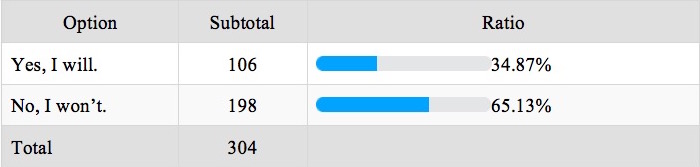}\caption{The answer result for question 3} \label{fig:fra}\end{figure}
\end{enumerate}

\subsection{The Distance Adjacency Matrix of the Map}
\begin{figure}[htbp]\centering\includegraphics[width=1\textwidth]{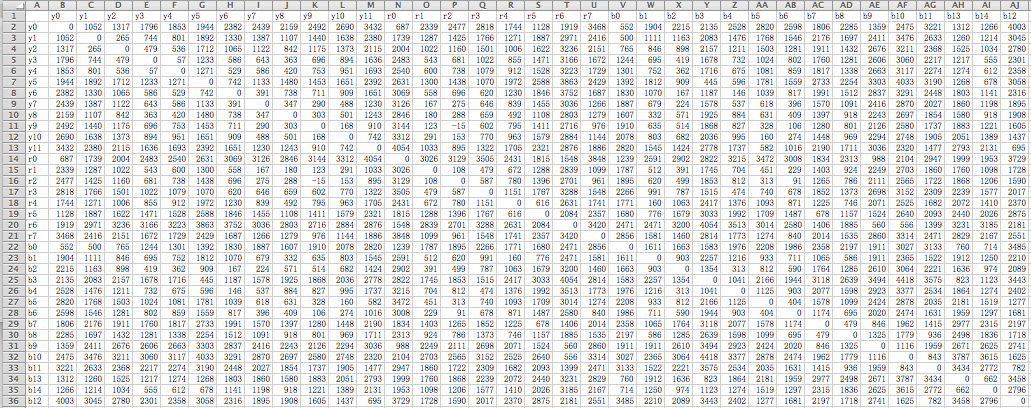}\caption{The Distance Adjacency Matrix of the Map} \label{fig:fra}\end{figure}

\subsection{Genetic Annealing}
\begin{lstlisting}[language=python]
\# 
\# Genetic.py
\# 
\# Copyright (c) Hanyi Luo, Mengzhan Liufu, Dongrong Li. All rights reserved
\# 
import json
import random

from matplotlib import pyplot
import pandas
import os

def simulate(p, c, m):
    global deliverymen
    data = []
    deliverymen = init(c)
    for t in range(len(order_time)):
        distribute_orders(t, m, p)
        for d in deliverymen:
            d.program()
        for i in range(cycle):
            for d in deliverymen:
                d.run(cycle * t + i)
    for d in deliverymen:
        for o in d.finished_orders:
            data.append({'time': o.start_time, 'avg_price': m, 'proportion': p, 'deliverymen_number': c, 'latency': o.latency})
        for o in d.orders:
            data.append({'time': o.start_time, 'avg_price': m, 'proportion': p, 'deliverymen_number': c, 'latency': -1})
    return pandas.DataFrame(data)

def distance(position, vertex):
    l = g[position[0]][position[1]]
    a = l * position[2]
    b = l * (1 - position[2])
    return (position[0], a, a + dist[position[0]][vertex]) if a + dist[position[0]][vertex] < b + dist[position[1]][vertex] else (position[1], b, b + dist[position[1]][vertex])

def argmin(array):
    return array.index(min(array))

def randSwap(new_route, successor):
    n = len(new_route)
    i = random.randint(0, n - 2)
    j = i + 1
    while new_route[i] in successor and new_route[j] in successor[new_route[i]]:
        i = random.randint(0, n - 2)
        j = i + 1
    new_route[i], new_route[j] = new_route[j], new_route[i]

def route_loss(position, route, tips):
    s = 0
    distSoFar = distance(position, route[0])[2]
    s += distSoFar * 2 ** (tips[route[0]] / MAX_TIP)
    for i in range(1, len(route)):
        distSoFar += dist[route[i-1]][route[i]]
        s += distSoFar * 2 ** (tips[route[i]] / MAX_TIP)
        if tips[route[i]] > 0 and distSoFar > 7500:
            return float('inf')
    return s

def program_route(position, orders):
    new_route = []
    route = []
    successor = {}
    tips = {}
    for order in orders:
        if order.status == 0:
            new_route.append(order.destination)
            if order.source in successor:
                successor[order.source].append(order.destination)
                tips[order.source] += order.price
            else:
                successor[order.source] = [order.destination]
                tips[order.source] = order.price
        else:
            route.append(order.destination)
        if order.destination in tips:
            tips[order.destination] += order.price
        else:
            tips[order.destination] = order.price
    route = list(set(route)) + list(successor.keys()) + list(set(new_route))
    losses = [route_loss(position, route, tips)]
    for _ in range(ITERATIONS):
        new_route = route[:]
        for i in range(int(temperature)+1):
            randSwap(new_route, successor)
        if random.random() < temperature \% 1:
            randSwap(new_route, successor)
        loss = route_loss(position, new_route, tips)
        delta = loss - losses[-1]
        if loss < losses[-1]:
            route = new_route
            losses.append(loss)
        else:
            losses.append(losses[-1])
    return route
 \end{lstlisting}

\end{document}